\documentstyle[12pt]{article}\textheight 220mm\textwidth 160mm

\newcommand{\al}{\alpha}	\newcommand{\be}{\beta}
\newcommand{\g}{\gamma}		\newcommand{\de}{\delta}
\newcommand{\e}{\epsilon}	
\newcommand{\z}{\zeta}		\newcommand{\th}{\theta}
	
\newcommand{\k}{\kappa}		\newcommand{\lam}{\lambda}
\newcommand{\m}{\mu}		\newcommand{\n}{\nu}
\newcommand{\x}{\xi}		\newcommand{\p}{\pi}
\newcommand{\r}{\rho}		\newcommand{\s}{\sigma}
\newcommand{\ta}{\tau}		
\newcommand{\vp}{\varphi}	\newcommand{\w}{\omega}
\newcommand{\h}{\eta}

		\newcommand{\Lam}{\Lambda}

\newcommand{\Th}{\Theta}

\newcommand{\dl}{\partial}

\newcommand{\nn}{\nonumber}
\newcommand{\CA}{{\cal A}}
\newcommand{\CD}{{\cal D}}
\newcommand{\CJ}{{\cal J}}
\newcommand{\CL}{{\cal L}}
\newcommand{\CQ}{{\cal Q}}
\newcommand{\CS}{{\cal S}}
\newcommand{\diag}{{\rm diag}}

\begin{document}

\hfill IU-MSTP/13

\hfill KANAZAWA-96-07

\hfill May, 1996
\begin{center}
 {\Large\bf Super-Virasoro Anomaly, Super-Weyl Anomaly\\
and the Super-Liouville Action for 2D Supergravity}
\end{center}

\vspace*{1cm}
\def\thefootnote{\fnsymbol{footnote}}
\begin{center}{\sc Takanori Fujiwara,}$^1$
{\sc Hiroshi Igarashi}$^2$
and  {\sc Tadao Suzuki}$^3$
\end{center}
\vspace*{0.2cm}
\begin{center}
{\em $\ ^{1}$ Department of Physics, Ibaraki University,
Mito 310, Japan}\\
{\em $\ ^{2}$ Graduate School of Science and Engineering,
Ibaraki University, Mito 310, Japan}\\
{\em $\ ^{3}$ Graduate School of Science and Technology, 
Kanazawa Universiy, \\
Kanazawa 920-11, Japan}
\end{center}
\vfill
\begin{center}
{\large\sc Abstract}
\end{center}
\noindent
The relation between super-Virasoro anomaly and super-Weyl anomaly 
in $N=1$ NSR superstring coupled with 2D supergravity is investigated 
from canonical theoretical view point. The WZW action canceling the 
super-Virasoro anomaly is explicitly constructed. It is super-Weyl 
invariant but nonlocal functional of 2D supergravity. The nonlocality 
can be remedied by the super-Liouvlle action, which in turn recovers the 
super-Weyl anomaly. The final gravitational effective action turns out 
to be local but noncovariant super-Liouville action, describing  
the dynamical behavior of the super-Liouville fields. The BRST invariance 
of this approach is examined in the superconformal gauge and in the 
light-cone gauge. 

\newpage
\pagestyle{plain}
\section{Introduction}

The characteristic feature of string theories at subcritical 
dimensions is that the conformal degree of the world-sheet metric 
variables being decoupled from the theory due to the local Weyl 
invariance at the classical level comes into dynamical play 
throught the Weyl anomaly. As was shown by Polyakov in his classic 
paper \cite{pol2}, this leads to Liouville action in the conformal gauge. 
Motivated by his work, the Liouville quantum theory has been 
investigated extensively in refs. \cite{ct,arvis,m}. In the original 
functional 
approach of Polyakov there arose a difficulty in handling the 
path integral for the conformal mode due to the translation 
noninvariance of the functional measure. This led the authors of 
refs. \cite{pol1,kpz} to the analysis in the 
light-cone gauge. The conformal gauge was investigated by the authors 
of refs. \cite{d,dk}. They noted that the gravitational scaling dimensions 
can be reproduced also in this gauge by imposing the functional 
measure ansatz that the Jacobian of the transformation between the 
translation noninvariant measure to translation invariant one is an 
exponential of a local action of Liouville type for the conformal 
mode. Their ansatz was examined by the heat kernel method in ref. 
\cite{mmh}. 

The conventional approaches to subcritical string theories and/or 2D 
quantum gravity more or less rely on path integral formalism. It is 
certainly desired to have a better understanding of the path 
integral results from a consistent canonical view point. In ref. 
\cite{fikt}, it was argued by noting the connection between the Virasoro 
anomaly and the Weyl anomaly that the Liouville action must be 
introduced as the Wess-Zumino-Witten (WZW) term to cancel the 
Virasoro anomaly to recover the world-sheet reparametrization 
invariance also in the canonincal treatment. This approach turned 
out to reproduce the primary results both for the light-cone and the 
conformal gauges. The purpose of the present paper is to extend 
the work done for bosonic string to fermionic theory \cite{ns,dz}. 
2D quantum supergravity coupled to superconformal matter has been 
investigated in the light-cone gauge in refs. \cite{gx,pz,aaz} 
and in the superconformal gauge in refs. \cite{dk,dhk}. Analyses 
based on the BRST formalism have been carried out in refs. 
\cite{kura,ih}. The connection between the (super-)Virasoro anomaly 
and the (super-)Weyl anomaly has been extensively studied in 
refs. \cite{gm,fikm,fikkt}. Formulation of 2D (super)gravity as anomalous 
gauge theory has been argued in refs. \cite{fik1,fir-kt}, where 
the (super-)Weyl anomaly is canceled by introducing additional 
degrees of freedom identified with the (super-)Liouville mode. 
The present work will give another exposition to the subject 
taken up in ref. \cite{fir-kt}. 

This paper is organized as follows. In Section 2, we will describe 
the super-Virasoro anomaly in canonical formalism and examine the 
local invariances of the string action. The super-Virasoro anomaly 
will be related to the anomaly in the covariant conservations of 
the stress tensor and the supercurrent, leading to the anomaly equations. 
In Section 3, we will examine the integrability of the super-Virasoro 
anomaly 
and solve the anomaly equations by noting their covariance under 
reparametrizations and local supersymmetry. The anomaly 
canceling WZW action is also constructed explicitly. Superfield formulation 
of the super-Virasoro anomaly is argued in Section 4 to make clear the 
geometrical meaning of the super-Virasoro anomaly and the WZW action. 
The counterterm derived there is nonlocal in the 2D supergravity variables. 
We will describe in Section 5 the cancellation of the nonlocality by the 
nonlocal covariant super-Liouville action. Quantization 
of 2D supergravity in superconformal gauge and light-cone gauge will be 
argued in Section 6 to see how the primary results found in the literature 
are reproduced in our approach. Our main concern there is to examine the 
BRST invariance. Section 7 is devoted to summary and discussion. We also 
provide some appendices. Appendix A deals with the summary of covariant 
BRST transformations. The super-Virasoro algebra with the cosmological 
term included is treated in Appendix B. The BRST gauge-fixing procedure 
is exposed in Appendix C in some detail in the case of light-cone gauge. 

\section{Super-Virasoro constraints in 2D superstring}
\setcounter{equation}{0}

The fermionic string of Neveu-Schwarz-Ramond can be formulated as 
2D supergravity described by the action \cite{dz}
\begin{equation}
S_X=-\int d^2x~e\biggl[~\frac{1}{2}
\bigl(g^{\al\be}\dl_\al X\dl_\be X
-i\overline\psi\r^\al\nabla_\al\psi\bigr)
+\overline\chi_\al\r^\be\r^\al\psi\dl_\be X
+\frac{1}{4}\overline\psi\psi\overline\chi_\al\r^\be\r^\al
\chi_\be~\biggr]~,
\label{str}
\end{equation}
where $X^\m$ and $\psi^\m$ ($\m=0,\cdots,D-1$) are, respectively, 
the bosonic and fermionic coordinates of string variables. The 
zweibein and the gravitino on the world-sheet are denoted by $e_\al{}^a$ 
and 
$\chi_\al$, respectively.\footnote{We choose $\h^{ab}=\diag(-1,1)$ and 
$\h^{\mu\nu}=\diag(-1,1,\cdots,1)$ for flat metrices. The world-sheet 
coordinates are denoted by $x^\al=(\tau,\s)$ for $\al=0,1$ and are 
assumed to take $-\infty<\s<+\infty$. It is straightforward to make 
the analysis on a finite interval of $\s$ so as to impose the 
Neveu-Schwarz or Ramond boudary conditions. We will use the notation 
$\dot A=\dl_\tau A$ and $A'=\dl_\s A$ for derivatives. Dirac 
matrices $\r^a$ ($a=0,1$) are chosen to be 
$\r^0=\s_2$, $\r^1=i\s_1$, and $\r^5\equiv\r^0\r^1=\s_3$, where $\s_k$ 
($k=1,2,3$) are Pauli matrices.} 

The classical action (\ref{str}) is invariant 
under the reparametrizations and local supersymmetry on the world-sheet. 
It also possesses the invariances under local Lorentz transformations, 
Weyl rescalings and fermionic symmetry. The latter symmetries can be 
used to eliminate some degrees of freedom of $e_\al{}^a$ and $\chi_\al$ 
from (\ref{str}). So it is convenient to define variables for zweibein by
\begin{equation}
\lam^\pm=\pm\frac{e_0{}^\pm}{e_1{}^\pm}~, \qquad\x=\ln(- e_1{}^+e_1{}^-)~, 
\qquad l=\frac{1}{2}\ln\biggl(-\frac{e_1{}^+}{e_1{}^-}\biggr)~, 
\label{zweibein}
\end{equation}
where $e_\al{}^\pm=e_\al{}^0 \pm e_\al{}^1$ \cite{fikkt,fir-kt}. 
As for the fermionic variables, we introduce $\psi_\pm$ by
\begin{equation}
\psi=\pmatrix{(-e_1{}^-)^{-\frac{1}{2}}\psi_-\cr
(e_1{}^+)^{-\frac{1}{2}}\psi_+}~,
\label{fermion}
\end{equation}
and define the gravitino fields by
\begin{equation}
\n_\pm=\frac{\chi_{0\pm}\pm\lam^\mp\chi_{1\pm}}{\sqrt{\mp e_1{}^\mp}}~, 
\qquad
\Lam_\pm=\frac{4\chi_{1\mp}}{\sqrt{\pm e_1{}^\pm}}~, \label{gravitino}
\end{equation}
where $\chi_{\al\mp}$ stands for the upper and lower components 
of $\chi_\al$.
 
The superconformal gauge $e_\al{}^a=\sqrt{e}\de_\al^a$, 
$\displaystyle{\chi_\al
=-\frac{1}{2}\r_\al\r^\be\chi_\be}$ corresponds to the conditions
\begin{equation}
\lam^\pm=1~, \qquad \nu_\mp=0~, \qquad l=0 ~. \label{scg}
\end{equation}

Under the local Lorentz transformations $\de e_\al{}^\pm=\pm\lam
e_\al{}^\pm$, the Weyl rescaling $\de e_\al{}^\pm=\phi e_\al{}^\pm$ 
and the fermionic symmetry $\de\chi_\al=i\r_\al\h$, $X^\m$, $\psi_\pm$, 
$\lam^\pm$ and $\n_\pm$ are invariant, while $\x$, $l$ and $\Lam_\pm$ are 
transformed as 
\begin{equation}
\de\x=2\phi~, \qquad \de l=\lam~,\qquad
\de\Lam_\pm=-4\h_\pm~, \label{ltransf}
\end{equation}
where $\h$ is an arbitrary Majorana spinor and $\h_\pm$ are the 
rescaled components of $\h$ as $\psi_\pm$ defined by (\ref{fermion}). 
In terms of these variables 
(\ref{str}) can be written as 
\begin{eqnarray}
S_X&=&\int d^2x\biggl[~\frac{(\dot X-\lam^+X')
(\dot X+\lam^-X')}{\lam^++\lam^-}
+\frac{i}{2}\psi_+(\dot\psi_+-\lam^+\psi_+')
+\frac{i}{2}\psi_-(\dot\psi_-+\lam^-\psi_-') \nonumber\\
&&+\frac{2}{\lam^++\lam^-}\bigl\{i(\dot X-\lam^+ X')\psi_-\n_+
-i(\dot X+\lam^-X')\psi_+\n_-+\psi_+\psi_-\n_+\n_- ~\bigr\}\biggr]~. 
\label{stract}
\end{eqnarray}
Since (\ref{str}) is invariant under (\ref{ltransf}), (\ref{stract}) does 
not contain $\x$, $l$ and $\Lam_\pm$. 

Let us denote by $P_\mu$ the canonical momentum conjugate to $X^\mu$
and assume the following set of fundamental super-Poisson brackets among
$X$, $P$ and $\psi_\pm$
\begin{equation}
\{X^\mu(\s)~,P_\nu(\s')\}=\de^\mu_\nu\de(\s-\s')~, \qquad
\{\psi^\mu_\pm(\s)~,\psi^\nu_\pm(\s')\}=-i\h^{\mu\nu}\de(\s-\s')~. 
\label{fsPb}
\end{equation}
Then the canonical theory of (\ref{str}) is characterized by the 
quantities defined by
\begin{equation}
\vp_\pm=\frac{1}{4}(P\pm X')^2\pm\frac{i}{2}\psi_\pm\psi_\pm'~, \qquad 
\CJ_\pm=\psi_\pm(P\pm X')~. \label{sVconst}
\end{equation}
They satisfy the classical super-Virasoro algebra
\begin{eqnarray}
&&\{\vp_\pm(\s),\vp_\pm(\s')\}
=\pm(\vp_\pm(\s)+\vp_\pm(\s'))\dl_\s\de(\s-\s')~, \nonumber \\
&&\{\CJ_\pm(\s),\vp_\pm(\s')\}=\pm\frac{3}{2}\CJ_\pm\dl_\s\de(\s-\s')
\pm\CJ'_\pm(\s)\de(\s-\s')~, \label{csV} \\
&&\{\CJ_\pm(\s),\CJ_\pm(\s')\}=-4i\vp_\pm(\s)\de(\s-\s')~, \nonumber
\end{eqnarray}
and generate fixed $\ta$ reparametrizations and supertransformations 
on the canonical variables $X$, $P$ and $\psi_\pm$. 
If we consider $e_\al{}^a$ and $\chi_\al$ as dynamical variables, 
(\ref{sVconst}) appears as super-Virasoro constraints among canonical 
variables. This can be seen directly from the fact that these 
quantities can also be obtained from (\ref{stract}) by taking the 
variations with respect to $\lam^\pm$ and $\nu_\mp$, i.e.,
\begin{equation}
\frac{\de S_X}{\de \lam^\pm}=-\vp_\pm~,\qquad
\frac{\de S_X}{\de \nu_\mp}=\pm i\CJ_\pm~. 
\label{sVconst 2}
\end{equation}
For most part of this paper, however, we will regard $e_\al{}^a$ and 
$\chi_\al$ as classical background fields and do not consider 
(\ref{sVconst}) as constraints.\footnote{We refer to (\ref{sVconst}) as 
super-Virasoro constraints for simplicity.} 

The canonical hamiltonian can be expressed as a linear combination of 
the constraints 
\begin{equation}
H_0=\int d\s[~\lam^+\vp_++\lam^-\vp_--i\nu_-\CJ_++i\nu_+\CJ_-~]~. 
\label{cHam}
\end{equation}
Then the super-Virasoro constraints (\ref{sVconst}) are subject to the 
equations of motion
\begin{eqnarray}
&&\dot \vp_\pm\mp\lam^\pm\vp_\pm'\mp2\lam^{\pm\prime}\vp_\pm
+\frac{3}{2}i\nu_\mp'\CJ_\pm+\frac{i}{2}\nu_\mp\CJ_\pm'=0 ~,
\label{classeqphi}\\
&&\dot\CJ_\pm\mp\lam^\pm\CJ_\pm'\mp\frac{3}{2}\lam^{\pm\prime}\CJ_\pm
\mp4\nu_\mp\vp_\pm=0~. \label{classeqJ}
\end{eqnarray}

The symmetric stress tensor $T_{\al\be}$ and the supercurrent $J^\al_\pm$ 
are defined by 
\begin{equation}
T_{\al\be}=-\frac{2}{e}\frac{\de S_X}{\de g^{\al\be}}~, \qquad
J^\al_\pm=-\frac{1}{2e}\frac{\de S_X}{\de \chi_{\al\mp}}~.
\label{stress tensor}
\end{equation}
Each component of $T_{\al\be}$ and $J^\al_\pm$ can be expressed 
as linear combinations of the constraints (\ref{sVconst}) as
\begin{eqnarray}
&&T_{00}=(\lam^+)^2(\vp_++\frac{i}{4}\Lam_+\CJ_+)
+(\lam^-)^2(\vp_-+\frac{i}{4}\Lam_-\CJ_-)~, \nonumber\\
&&T_{01}=T_{10}=\lam^+(\vp_++\frac{i}{4}\Lam_+\CJ_+)
-\lam^-(\vp_-+\frac{i}{4}\Lam_-\CJ_- )~, \nonumber\\
&&T_{11}=\vp_++\frac{i}{4}\Lam_+\CJ_+
+\vp_-+\frac{i}{4}\Lam_-\CJ_-~, \nonumber\\
&&J^0_\pm=\frac{1}{2e}\CJ_\pm~, \qquad
J^1_\pm=\mp\frac{\lam^\pm}{2e}\CJ_\pm~.\label{components}
\end{eqnarray}
We again see that the stress tensor and supercurrent constraints 
$T_{\al\be}=\CJ^\al_\pm=0$ are fulfilled by imposing the super-Virasoro 
constraints $\vp_\pm=\CJ_\pm=0$ for $e_\al{}^a$ and $\chi_\al$ being 
dynamical variables. 

{}From the expressions (\ref{components}) it is easy to see that 
the traces of the stress tensor and the supercurrent vanish, i.e.,
\begin{equation}
T^\al_\al=0~, \qquad \r_\al J^\al=0~, \label{supertrace}
\end{equation}
where $J^\al$ is the two component spinor with $J^\al_\mp$ as 
upper and lower components, and use has been made of the relations 
$g^{00}(\lam^\pm)^2\pm g^{01}\lam^\pm+g^{11}=0$, 
$\r^\al\r^\be\r_\al=0$. Eq. (\ref{supertrace}) is the direct 
consequence of the super-Weyl invariance of (\ref{str}).
The invariances of (\ref{str}) under reparametrizations and 
local supersymmetry can be restated by a kind of generalization 
of covariant conservations of stress tensor and supercurrent. 

To see this let us denote by $\de_u$ the Lie derivative for 
the reparametrization $x^\al\rightarrow x^\al+u^\al(x)$, then 
$\lam^\pm$ and $\nu_\mp$ are transformed by
\begin{eqnarray}
\de_u\lam^\pm&=&-u^\al\partial_\al\lam^\pm
-\lam^\pm(\dot u^0\mp\lam^\pm u^{0\prime})
\mp(\dot u^1\mp\lam^\pm u^{1\prime})~, \nonumber\\
\de_u\nu_\mp&=&-u^\al\partial_\al\nu_\mp
-\Bigl\{\dot u^0\mp\lam^\pm u^{0\prime}-\frac{1}{2}
(u^{1\prime}\pm\lam^\pm u^{0\prime})\Bigr\}\nu_\mp~. \label{dunu}
\end{eqnarray}
The infinitesimal supertransformations are given by
\begin{eqnarray}
\de_\e\lam^\pm&=&-4i\e_\mp\nu_\mp~, \nonumber\\
\de_\e\nu_\mp&=&\dot\e_\mp\mp\lam^\pm\e_\mp'
\pm\frac{1}{2}\lam^{\pm\prime}\e_\mp~, \label{localsusy}
\end{eqnarray}
where the parameters $\e_\mp$ of local supersymmetry are the 
rescaled components of a two component Majorana spinor $\e$ as 
is defined in (\ref{fermion}) for $\psi$. 
The variations of (\ref{stract}) under these transformations are 
computed to 
\begin{eqnarray}
\de_uS_X&=&\int d^2x\biggl(~\de_u\lam^+
\frac{\de}{\de \lam^+}
+\de_u\lam^-\frac{\de}{\de \lam^-}
+\de_u\nu_-\frac{\de}{\de \nu_-}
+\de_u\nu_+\frac{\de}{\de \nu_+}~\biggr)S_X
\nonumber\\
&=&\int d^2x\biggl(~-(u^1+\lam^+u^0)
(\dot \vp_+-\lam^+\vp_+'-2\lam^{+\prime}\vp_+
+\frac{3}{2}i\nu_-'\CJ_++\frac{i}{2}\nu_-\CJ_+') \nonumber\\
&&~~~~~~~~+(u^1-\lam^-u^0)
(\dot \vp_-+\lam^-\vp_-'+2\lam^{-\prime}\vp_-
+\frac{3}{2}i\nu_+'\CJ_-+\frac{i}{2}\nu_+\CJ_-') \nonumber\\
&&~~~~~~~~+iu^0\nu_-
(\dot\CJ_+-\lam^+\CJ_+'-\frac{3}{2}\lam^{+\prime}\CJ_+
-4\nu_-\vp_+) \nonumber\\
&&~~~~~~~~-iu^0\nu_+
(\dot\CJ_-+\lam^-\CJ_-'+\frac{3}{2}\lam^{-\prime}\CJ_-
+4\nu_+\vp_-)~\biggr)~, \label{covconsvp} \\
\de_\e S_X&=&\int d^2x\biggl(~\de_\e\lam^+
\frac{\de}{\de \lam^+}
+\de_\e\lam^-\frac{\de}{\de \lam^-}
+\de_\e\nu_-\frac{\de}{\de \nu_-}
+\de_\e\nu_+\frac{\de}{\de \nu_+}~\biggr)S_X
\nonumber\\
&=&\int d^2x\biggl(~-i\e_-
(\dot\CJ_+-\lam^+\CJ_+'-\frac{3}{2}\lam^{+\prime}\CJ_+
-4\nu_-\vp_+) \nonumber\\
&&~~~~~~~~~~+i\e_+(\dot\CJ_-+\lam^-\CJ_-'+\frac{3}{2}\lam^{-\prime}\CJ_-
+4\nu_+\vp_-)~\biggr)~, \label{covconsJ}
\end{eqnarray}
where use has been made of the equations of motion for $X$ and 
$\psi_\pm$. The local invariances of (\ref{str}) thus lead to the 
equaitons of motion (\ref{classeqphi}) and (\ref{classeqJ}). 

We now consider canonical quantization of string variables $X$, 
$P$ and $\psi_\pm$ by the quantization rule $\{~,~\}\rightarrow
-i[~,~]$ in (\ref{fsPb}), where $[~,~]$ stands for supercommutator.
The super-Virasoro constraints (\ref{sVconst}) become quantum 
mechanical operators. They are, however, ill-defined unless an 
operator ordering is specified. One way to implement this is 
to Fourier expand field variables and then to define harmonic 
oscillators. By putting rasing operators to the left of 
lowering operators, we obtain the normal ordered form of an 
arbitrary product of operators of equal $\ta$. The operators 
$P\pm X'$ and $\psi_\pm$ can be divided into two parts, one 
containing only lowering operators, $(P\pm X')^{(+)}$ and 
$\psi_\pm^{(+)}$, and the other containing only rasing operators
$(P\pm X')^{(-)}$ and $\psi_\pm^{(-)}$. They are defined by
\begin{eqnarray}
(P+X')^{(\pm)}(\ta, \s)&=&\int d\s'\de^{(\mp)}(\s-\s')
(P+X')(\ta,\s')~, \nonumber\\
(P-X')^{(\pm)}(\ta, \s)&=&\int d\s'\de^{(\pm)}(\s-\s')
(P-X')(\ta,\s')~, \nonumber\\
\psi_+^{(\pm)}(\ta,\s)&=&\int d\s'\de^{(\mp)}(\s-\s')\psi_+(\ta,\s')~, 
\nn \\
\psi_-^{(\pm)}(\ta,\s)&=&\int 
d\s'\de^{(\pm)}(\s-\s')\psi_-(\ta,\s') ~, \label{creanniop}
\end{eqnarray}
with $\displaystyle{\de^{(\pm)}(\s)=\frac{1}{2\pi}\frac{\pm i}
{\s\pm i\e}}$. Then by putting $(P\pm X')^{(-)}$ and $\psi_\pm^{(-)}$ 
to the left of $(P\pm X')^{(+)}$ and $\psi_\pm^{(+)}$, we can define 
an operator ordering in the present case. 

The super-Virasoro operator thus defined will develop super-Virasoro 
anomaly in their supercommutation relations, i.e.,
\begin{eqnarray}
&&[\vp_\pm(\s),\vp_\pm(\s')]
=\pm i(\vp_\pm(\s)+\vp_\pm(\s'))\dl_\s\de(\s-\s')
\pm i\k_0\partial_\s^3\de(\s-\s')~, \nonumber \\
&&[\CJ_\pm(\s),\vp_\pm(\s')]=\pm i\frac{3}{2}\CJ_\pm\dl_\s\de(\s-\s')
\pm i\CJ'_\pm(\s)\de(\s-\s')~, \label{qsV} \\
&&[\CJ_\pm(\s),\CJ_\pm(\s')]=4\vp_\pm(\s)\de(\s-\s')
+8\k_0\partial_\s^2\de(\s-\s')~ \nonumber
\end{eqnarray}
where $\k_0$ is given by
\begin{equation}
\k_0=-\frac{D}{16\pi}~. \label{mcc}
\end{equation}
Due to the appearance of the super-Virasoro anomaly, the classical 
equations of motion (\ref{classeqphi}) and (\ref{classeqJ}) are 
modified to 
\begin{eqnarray}
&&\dot \vp_\pm\mp\lam^\pm\vp_\pm'\mp2\lam^{\pm\prime}\vp_\pm
+\frac{3}{2}i\nu_\mp'\CJ_\pm+\frac{i}{2}\nu_\mp\CJ_\pm'
=\pm \k_0\lam^{\pm\prime\prime\prime} ~,
\label{qeqphi}\\
&&\dot\CJ_\pm\mp\lam^\pm\CJ_\pm'\mp\frac{3}{2}\lam^{\pm\prime}\CJ_\pm
\mp4\nu_\mp\vp_\pm
=\pm 8\k_0\nu_\mp''~. \label{qeqJ}
\end{eqnarray}
Putting these into (\ref{covconsvp}) and (\ref{covconsJ}), we obtain
\begin{eqnarray}
\de_uS_X&=&-\k_0\int d^2x\bigl(~(u^1+\lam^+u^0)
\lam^{+\prime\prime\prime}+(u^1-\lam^-u^0)\lam^{-\prime\prime\prime} 
\nonumber\\
&&~~~~~~~~~~~~~~~~~~~-8iu^0(\nu_-\nu_-''+\nu_+\nu_+'')~\bigr)~, 
\label{qcovconsvp} \\
\de_\e S_X&=&-8i\k_0\int d^2x\bigl(~\e_-\nu_-''
+\e_+\nu_+''~\bigr)~. \label{qcovconsJ}
\end{eqnarray}
We see that invariances of (\ref{str}) under the reparametrizations 
and local supertransformations are violated by the super-Virasoro 
anomaly unless the background 2D metric variables satisfy 
\begin{equation}
\partial_\s^3\lam^\pm=0 ~, \qquad
\partial_\s^2\nu_\mp=0 ~. \label{afbgf}
\end{equation}
The superconformal gauge (\ref{scg}) is a special case satisfying 
these condition. 

The reason for the noninvariance of $S_X$ is rather 
obvious. The ordering prescription we are employing to define operator 
products involves equal $\ta$ operators of different spatial 
coordinates and, hence, treats space- and time-coordinates 
asymmetrically. This violates the supercovariance on the world-sheet.
Since we do not refer to any particular zweibein $e_{\al}{}^a$ and 
gravitino $\chi_\al$ in defining ordering prescription, the super-Weyl 
invariance remains intact upon quantization. At first sight, this 
conclusion seemed to be inconsistent with the well-known fact that the 
reparametrization invariance and local supersymmetry can be 
maintained upon quantization, while the super-Weyl invariance is 
violated by the super-Weyl anomaly. In Section 5
we will resolve this puzzling feature.

Before closing this section it is worth mentioning the survival 
symmetries of (\ref{str}) at the quantum level. From (\ref{qcovconsvp}) 
and (\ref{qcovconsJ}), $S_X$ is invariant for
\begin{equation}
\partial_\s u^0=\partial_\s^3u^1=0~, \qquad
\partial_\s^2\e_\mp=0~. \label{gsl2R}
\end{equation}
As we will see in Section 4, (\ref{gsl2R}) can be enlarged 
corresponding to the ten anomaly free components of the 
super-Virasoro generators.

\section{Super-Virasoro anomaly}\setcounter{equation}{0}

In the previous section we have argued that the super-Virasoro 
anomaly originates from the artificial definition of operator 
ordering. It is known, however, that (\ref{str}) possesses no 
anomalies under the reparametrizations and local supersymmetry. 
This implies that we can find a suitable redefinition of (\ref{str}) 
to maintain these invariances at the quantum level. In this section 
we shall see that the super-Virasoro anomaly can be canceled by 
adding to (\ref{str}) a counterterm, which is a functional of 
$\lam^\pm$ and $\nu_\mp$. 

The super-Virasoro anomaly in the equations of motion for the 
constraints (\ref{qeqphi}) and (\ref{qeqJ}) can be canceled 
if we modify the constraints (\ref{sVconst}) by
\begin{equation}
\bar\vp_\pm=\vp_\pm+\z_\pm~, \qquad 
\bar\CJ_\pm=\CJ_\pm+j_\pm ~,
\label{msV}
\end{equation}
where $\z_\pm$ and $j_\pm$ are the solutions to the equations 
\begin{eqnarray}
&&\dot \z_\pm\mp\lam^\pm\z_\pm'\mp2\lam^{\pm\prime}\z_\pm
+\frac{3}{2}i\nu_\mp'j_\pm+\frac{i}{2}\nu_\mp j_\pm'
=\mp \k_0\lam^{\pm\prime\prime\prime}~,
\label{eqz}\\
&&\dot j_\pm\mp\lam^\pm j_\pm'\mp\frac{3}{2}\lam^{\pm\prime} j_\pm
\mp4\nu_\mp\z_\pm
=\mp8\k_0\nu_\mp''~. \label{eqj}
\end{eqnarray}
They depend functionally on $\lam^\pm$, $\nu_\mp$ and are assumed 
to vanish if there is no super-Virasoro anomaly in (\ref{qeqphi}) 
and (\ref{qeqJ}), i.e., 
\begin{equation}
\z_\pm=j_\pm=0 \quad {\rm for} \quad \lam^\pm{}'''=\nu_\mp''=0~. 
\label{BC}
\end{equation}

The modification in (\ref{msV}) can be readily related to the 
counterterm canceling the super-Virasoro anomaly (\ref{qcovconsvp}) 
and (\ref{qcovconsJ}). Let us denote the counterterm by $S_V$, then 
$\z_\pm$ and $j_\pm$ can be obtained by
\begin{equation}
\frac{\de S_V}{\de \lam^\pm}=-\z_\pm~,\qquad
\frac{\de S_V}{\de \nu_\mp}=\pm ij_\pm~,
\label{Sv}
\end{equation}
as is inferred form (\ref{sVconst 2}). The $S_V$,  is assumed to be a 
functional of $\lam^\pm$ and $\nu_\mp$. Eq. (\ref{Sv}) can be regarded 
as functional differential equations. In order for them to be 
integrable the super-Virasoro anomaly must satisfy the Wess-Zumino 
consistency conditions \cite{wz}. (See also \cite{tanii,gm}.)
To see this let us define the generators 
of time-preserving reparametrization and local supersymmetry 
$\CL^\pm_u$ and $\CQ^\pm_\e$ by
\begin{eqnarray}
\CL^\pm_u&=&\int d^2x\biggl(~\de_u\lam^\pm\frac{\de}{\de\lam^\pm}
+\de_u\nu_\mp\frac{\de}{\de\nu_\mp}~\biggr)~, \label{Lu}\\
\CQ^\pm_\e&=&\int d^2x\biggl(~\de_\e\lam^\pm\frac{\de}{\de\lam^\pm}
+\de_\e\nu_\mp\frac{\de}{\de\nu_\mp}~\biggr)~, \label{Qe}
\end{eqnarray}
where $\de_\e$ is given by (\ref{localsusy}) and $\de_u$ is obtained 
from (\ref{dunu}) for $u^0(x)=0$ and $u^1(x)=u(x)$, i.e.,
\begin{equation}
\de_u\lam^\pm=\mp\dot u+\lam^\pm u'-u\lam^\pm{}'~, \qquad
\de_u\nu_\mp=-u\nu_\mp'+\frac{1}{2}u'\nu_\mp~. \label{tpLd}
\end{equation}
These generators satisfy the classical super-Virasoro algebra 
\begin{equation}
[\CL^\pm_u,\CL_v^\pm]=\CL_{[u,v]}^\pm~, \qquad
[\CL^\pm_u,\CQ^\pm_\e]=\CQ^\pm_{u\e'-\frac{1}{2}u'\e}~, \qquad
[\CQ^\pm_{\e_1}, \CQ^\pm_{\e_2}]=4\CL^\pm_{\pm i\e_{1\mp}\e_{2\mp}}~, 
\label{csVa}
\end{equation}
where we have defined $[u,v]=uv'-vu'$.
Then (\ref{eqz}) and (\ref{eqj}) can be written as 
\begin{equation}
\CL^\pm_uS_V=\CA^\pm_u~, \qquad \CQ^\pm_\e S_V=\CS^\pm_\e~, 
\label{aeq}
\end{equation}
where $\CA^\pm_u$ and $\CS^\pm_\e$ are the super-Virasoro anomaly 
given by
\begin{equation}
\CA^\pm_u=\k_0\int d^2x~u\lam^\pm{}'''~, \qquad 
\CS^\pm_\e=8i\k_0\int d^2x~\e_\mp\nu_\mp''~.
\label{sVan}
\end{equation}
From (\ref{csVa}) and (\ref{aeq}) we obtain the Wess-Zumino conditions 
\begin{eqnarray}
&&\CL^\pm_u\CA^\pm_v-\CL^\pm_v\CA^\pm_u=\CA_{[u,v]}^\pm~, \nonumber\\
&&\CL^\pm_u\CS^\pm_\e-\CQ^\pm_\e\CA_u^\pm
=\CS^\pm_{u\e'-\frac{1}{2}u'\e}~, \label{WZc} \\
&&\CQ^\pm_{\e_1}\CS^\pm_{\e_2}-\CQ^\pm_{\e_2}\CS^\pm_{\e_1}
=4\CA_{\mp i\e_{1\mp}\e_{2\mp}}^\pm~. \nonumber
\end{eqnarray}
Since (\ref{sVan}) satisfies these conditions, we see that (\ref{Sv})
is indeed integrable. 

The solution to (\ref{eqz}) and (\ref{eqj}) satisfying (\ref{BC})
can be obtained by utilizing the transformation properties of 
$\z_\pm$ and $j_\pm$ under time-preserving reparametrizations and 
local supertransformations. Let us first consider a time- and 
orientation-preserving transformation given by $x^0\rightarrow 
\tilde x^0=x^0$, $x^1\rightarrow \tilde x^1=f(x)$ with $f'(x)>0$. 
Under the coordinate change $\lam^\pm$ 
and $\nu_\mp$ are transformed into 
\begin{equation}
\tilde\lam^\pm(\tilde x)=\mp\dot f(x)+f'(x)\lam^\pm(x)~, \qquad
\tilde\nu_\mp(\tilde x)=\sqrt{f'(x)}\nu_\mp(x)~. \label{tpct} 
\end{equation}
In the new coordinates $(\tilde x)$ the same type of equations 
as (\ref{eqz}) and (\ref{eqj}) must be satisfied by 
$\tilde\z_\pm(\tilde x)$ and $\tilde j_\pm(\tilde x)$. This 
unambiguously determines transformation properties
\begin{equation}
\tilde\z_\pm(\tilde x)=(f'(x))^{-2}(\z_\pm(x)+\k_0\CD f(x))~, \qquad
\tilde j_\pm(\tilde x)=(f'(x))^{-\frac{3}{2}}j_\pm(x)~,
\label{tpctzj}
\end{equation}
where $\CD f$ is the schwarzian derivative defined by 
\begin{equation}
\CD f=\frac{f'''}{f'}-\frac{3}{2}\biggl(\frac{f''}{f'}\biggr)^2~.
\label{SchwD}
\end{equation}
We see that $\z_\pm$ and $j_\pm$, respectively, are tensors of 
weight $2$ and $3/2$. The inhomogeneous term denotes the finite 
form of Virasoro anomaly analogous to the one well-known in conformal 
field theories. 

We next consider the supertransformation (\ref{localsusy}). By 
requiring covariance of (\ref{eqz}) and (\ref{eqj}), we find that 
$\z_\pm$ and $j_\pm$ must be transformed by 
\begin{equation}
\de_\e\z_\pm=-\frac{i}{2}\e_\mp j_\pm'-\frac{3}{2}i\e_\mp'j_\pm~,
\qquad
\de_\e j_\pm=\pm4\e_\mp\z_\pm\mp8\k_0\e_\mp''~. \label{zjsusy}
\end{equation}
The term proportional to $\k_0$ represents anomalous behavior 
of $j_\pm$. The finite form of these infinitesimal transformations 
can be obtaind by straightforward integration. Let us denote by $\h_\mp$ 
the components of a finite Majorana spinor $\h$, then the 
supertransformation of $\lam^\pm$ and $\nu_\mp$ by $\h$ are given by
\begin{eqnarray}
\hat\lam^\pm&=&\lam^\pm-4i\h_\mp\nu_\mp
-2i\h_\mp(\dot\h_\mp\mp\lam^\pm\h_\mp')~, \nonumber\\
\hat\nu_\mp&=&\nu_\mp+\dot\h_\mp\mp\lam^\pm\h_\mp'
\pm\frac{1}{2}\lam^\pm{}'\h_\mp\mp3i\h_\mp\h_\mp'\nu_\mp
\mp i\h_\mp\h_\mp'\dot\h_\mp~. \label{fsusyln}
\end{eqnarray}
For example, $\hat \lam^+$ can be obtained by Taylor expanding 
$\lam^+(\al)\equiv e^{\al\CQ^+_\h}\lam^+e^{-\al\CQ^+_\h}$ with 
respect to $\al$ and then putting $\al=1$. Similarly, we find 
\begin{eqnarray}
\hat\z_\pm&=&\z_\pm-\frac{3}{2}i\h_\mp'j_\pm-\frac{i}{2}\h_\mp j'_\pm
\pm2i\h_\mp\h_\mp'\z_\pm\pm2i\k_0(\h_\mp\h_\mp'''+3\h_\mp'\h_\mp'') ~, 
\nonumber\\
\hat j_\pm&=&j_\pm\pm4\h_\mp\z_\pm\mp3i\h_\mp\h_\mp'j_\pm
\mp8\k_0(1\mp i\h_\mp\h_\mp')\h_\mp'' ~. \label{fsusyzj}
\end{eqnarray}
The $\k_0$-terms emerge from the anomalous transformation properties 
of $j_\pm$ and denote finite anomaly under local supertransformation. 

All these properties are enough to solve (\ref{eqz}) and (\ref{eqj}). 
We first choose $\h_\mp$ to satisfy $\hat\nu_\mp=0$. We next consider 
a coordinate transformation $\tilde x^0=x^0$, $\tilde x^1=f_\pm(x)$ 
where $\tilde\lam^\pm=\mp\dot f_\pm+f_\pm'\hat\lam^\pm=0$ is 
satisfied. Since $\tilde\nu_\mp$ also vanishes in this 
coordinates, (\ref{eqz}) and (\ref{eqj}) take the simplest forms
$\partial_{\tilde\ta}\tilde\z_\pm=\partial_{\tilde\ta}
\tilde j_\pm=0$ with obvious solution
\begin{equation}
\tilde\z_\pm=\tilde j_\pm=0~. \label{tzj}
\end{equation}
Since we know the transformation porperties of $\z_\pm$ and $j_\pm$ 
as (\ref{tpctzj}) and (\ref{fsusyzj}), we can completly determine 
$\z_\pm$ and $j_\pm$ from (\ref{tzj}) as
\begin{eqnarray}
\z_\pm&=&-\k_0(1\pm2i\h_\mp\h_\mp')\CD f_\pm
\pm2i\k_0(\h_\mp\h_\mp'''+3\h_\mp'\h_\mp'')~, \nonumber\\
j_\pm&=&\pm4\k_0\h_\mp\CD f_\pm\pm8\k_0(1\mp i\h_\mp\h_\mp')\h_\mp''~,
\label{solzj}
\end{eqnarray}
where $f_\pm$ and $\h_\mp$ are related to $\lam^\pm$ and $\nu_\mp$ 
by the conditions $\hat\nu_\mp=\tilde\lam^\pm=0$. Solving 
these conditions with respect to $\lam^\pm$ and $\nu_\mp$, we obtain
\begin{eqnarray}
\lam^\pm&=&(\pm1+2i\h_\mp\h_\mp')\frac{\dot f_\pm}{f_\pm'}
-2i\h_\mp\dot\h_\mp~, \nonumber\\
\nu_\mp&=&-(1\mp i\h_\mp\h_\mp')\dot\h_\mp
+\frac{\dot f_\pm}{f_\pm'}\h_\mp'
-\frac{1}{2}\biggl(\frac{\dot f_\pm}{f_\pm'}\biggr)'\h_\mp~.
\label{fpm}
\end{eqnarray}
Eqs. (\ref{solzj}) and (\ref{fpm}) are the extension of the results 
obtained for Polyakov string \cite{fikt} to superstring as one can 
easily see 
by putting $\h_\mp=0$. Since $f_\pm$ and $\h_\mp$ depend only on 
$\lam^\pm$ and $\nu_\mp$, they are invariant under the super-Weyl 
transformations. 

We are now in a position to compute the counterterm $S_V$ satisfying 
(\ref{Sv}). Since $\z_+$ and $j_+$ ($\z_-$ and $j_-$) depend only on 
$\lam^+$ and $\nu_-$ ($\lam^-$ and $\nu_+$), we can find a solution 
in the form 
\begin{equation}
S_V=S_V^++S_V^- ~, \label{ctSV}
\end{equation}
where $S_V^+$ ($S_V^-$) is a functional of $\lam^+$, $\nu_-$ 
($\lam^-$, $\nu_+$) and must satisfy the variational equation
\begin{equation}
\de S_V^\pm=\int d^2x(~-\de \lam^\pm\z_\pm\pm i\de \nu_\mp j_\pm~) ~.
\label{dSvpm}
\end{equation}
Putting (\ref{solzj}) into the rhs' of these expressions and using 
the relations (\ref{fpm}), we obtain the counterterms
\begin{eqnarray}
S_V^\pm&=&\pm \k_0\int d^2x
\biggl[~\frac{1}{2}\biggl\{\frac{(f_\pm'')^2\dot f_\pm}{(f_\pm')^3}
-\frac{f_\pm''\dot f_\pm'}{(f_\pm')^2}\biggr\}
\mp4i\dot\h_\mp\h_\mp'' \nonumber\\
&&~~~~~~~~~~~~~~~\pm2i(\h_\mp\h_\mp'''+3\h_\mp'\h_\mp'')
\frac{\dot f_\pm}{f_\pm'}
-2\dot\h_\mp\h_\mp\h_\mp'\h_\mp''~\biggr]. \label{Svpm}
\end{eqnarray}
This is the extension of the results for Polyakov 
string found in ref. \cite{fikt} to superstring. 

Since $S_V$ is manifestly invariant under super-Weyl rescalings, 
and cancels the super-Virasoro anomaly (\ref{qcovconsvp}) and 
(\ref{qcovconsJ}) in $S_X$ by construction, we see that 
the effective action $S_X+S_V$ possesses all the local 
symmetries of the classical theory and there is no anomaly in 
super-Weyl symmetry. This seemed again to be inconsistent with 
the generally 
accepted fact that one can not maintain reparametrization invariance 
and local supersymmetry at the quantum level without sacrificing 
super-Weyl invariance. As was noted in ref. \cite{fikt} for bosonic 
theory, this is not a 
contradiction. Since $f_\pm$ and $\h_\mp$ satisfying (\ref{fpm}) 
depend nonlocally on $\lam^\pm$ and $\nu_\mp$, the counterterm 
$S_V$ is in general a nonlocal functional of these variables. 
It is, however, possible to choose the counterterm recovering 
reparametrization invariance and supersymmetry to be a local 
functional of $e_\al{}^a$ and $\chi_\al$ as is known by 
perturbative analysis. We will in fact show in Section 5
that we can reproduce super-Weyl anomaly by requiring the locality 
of the counteraction. 

\section{Superspace formulation of the super-Virasoro anomaly}
\setcounter{equation}{0}

In the previous section we have obtained the counterterm $S_V$ that cancels 
the super-Virasoro anomaly to recover reparametrization invariance 
and local supersymmetry at the quantum level. In this section we will 
make a small detour to superspace and argue the superspace formulation 
of the super-Virasoro anomaly. We wish to note the existence of somewhat 
unusual fermionic coordinates and supertranslations under which $\lam^\pm$ 
and $\n_\mp$ form supermultiplets and to clarify the geometrical 
meaning of what we have done in the previous section. We also see that 
the arguments of ref. \cite{fikt} on the Virasoro anomaly in bosonic string 
can naturally extend to the supersymmetric case and the invariance of 
$S_V^\pm$ under the OSp(1,2) Kac-Moody like transformations reveals itself 
in the reformulation by superfields.

Let us consider superspaces with supercoordinates 
$z^\pm =( \tau,\sigma,\theta_\mp)$ for $\lam^\pm$ and $\n_\mp$, 
where $\th_\mp$ are grassmannian variables. The infinitesimal 
supertranslations for $z^\pm$ are defiend by
\begin{equation}
{\delta}_{\epsilon} \tau =0~,\qquad
{\delta}_{\epsilon} \sigma =\pm2i\epsilon_\mp \theta_\mp~,\qquad
{\delta}_{\epsilon} \theta_\mp =\epsilon_\mp ~.
\end{equation}
The supercovariant derivatives are given by 
\begin{equation}
D_\pm=\pm i\frac{\dl}{\dl \theta_\mp}+2\theta_\mp \frac{\dl}{\dl \sigma} ~.
\label{scovd}
\end{equation}
They anticommute with the supercharges and satisfy $D_\pm^2=\pm2i\dl_\s$.

We now define superfields by 
\begin{equation}
 \phi^\pm(z^\pm)=\lambda^\pm(x)+4i\theta_\mp\nu_\mp(x)~, \qquad
   G_\pm(z^\pm)=j_\pm(x)\mp4\theta_\mp\zeta_\pm(x)~. \label{superfields}
\end{equation}
Under the time-preserving supercoordinate transformations given by 
\begin{eqnarray}
z^\pm=(\tau,\s,\th_\mp)\rightarrow 
\overline z{}^\pm=(\tau,g^\pm(z^\pm),\vartheta_\mp(z^\pm)) \quad
{\rm with} \quad 
D_\pm g^\pm\pm2i\vartheta_\mp D_\pm\vartheta_\mp=0 ~, \label{tpsct}
\end{eqnarray}
(\ref{superfields}) are transformed into
\begin{eqnarray}
\overline\phi^\pm(\overline z^\pm)&=&(\mp iD_\pm\vartheta_\mp(z^\pm))^2
(\phi^\pm(z^\pm)\mp\dot g^\pm(z^\pm)+2i\vartheta_\mp(z^\pm)
\dot\vartheta_\mp(z^\pm)) ~, \nn\\
\overline G_\pm(\overline z^\pm)&=&(\mp iD_\pm\vartheta_\mp(z^\pm))^{-3}
(G_\pm(z^\pm)+2i\k_0{\cal D}_\pm\vartheta_\mp(z^\pm)) ~, \label{tps}
\end{eqnarray}
where ${\cal D}_\pm$ stand for superschwarzian derivatives defined by 
\begin{eqnarray}
{\cal D}_\pm\vartheta_\mp=\frac{D_\pm^4\vartheta_\mp}{D_\pm\vartheta_\mp}-2
\frac{D_\pm^2\vartheta_\mp D_\pm^3\vartheta_\mp}{(D_\pm\vartheta_\mp)^2} ~. 
\label{supschwd}
\end{eqnarray}
In terms of (\ref{superfields}), the anomaly equations (\ref{eqz}) 
and (\ref{eqj}) can be written as 
\begin{equation}
  \dot G_\pm  +\frac{i}{2} \phi^\pm D^2_\pm G_\pm 
+ \frac{i}{4} D_\pm\phi^\pm D_\pm G_\pm
+ \frac{3}{4} i D^2_\pm \phi^\pm G_\pm
= \frac{\kappa_0}{2} D^5_\pm \phi^\pm ~. \label{sfeq}
\end{equation}
The transformation properties for $G_\pm$ can be derived by noting 
the covariance of  (\ref{sfeq}) under (\ref{tpsct}). 

We can solve (\ref{sfeq}) by making use of the supercoordinate 
transformations which transform $\phi^\pm$ into 0. Let us denote such 
transformations by $z^\pm=(\tau,\s,\th_\mp) \rightarrow 
\overline z{}^\pm=(\tau,F^\pm(z^\pm),$ $\Th_\mp(z^\pm))$, then we find 
\begin{eqnarray}
\phi^\pm&=&(\mp iD_\pm\Th_\mp)^{-2}(\pm \dot F^\pm-2i\Th_\mp\dot\Th_\mp) ~, 
\nn\\
G_\pm&=&-2i\k_0{\cal D}_\pm\Th_\mp ~. \label{asol}
\end{eqnarray}
It is easy to see that $F^\pm$ and $\Th_\mp$ are transformed as scalar 
superfields under the supercoordinate changes (\ref{tpsct}). 
Eqs.(\ref{asol}) correspond to the solutions (\ref{fpm}) and (\ref{solzj}) 
for the choices 
\begin{eqnarray}
F^\pm=f_\pm\mp2if_\pm'\th_\mp\h_\mp ~,\qquad
\Th_\mp=\sqrt{f_\pm'}\{\h_\mp+\th_\mp(1\mp i\h_\mp\h_\mp{}')\} ~, 
\label{FTh}
\end{eqnarray}
where $f_\pm$ and $\h_\mp$ have been introduced in the previous section. 

We have furnished all the staffs needed to write down the counterterms 
(\ref{Svpm}) as functionals on the superspace. It is straightforward 
to retrace the procedure to obtain (\ref{Svpm}) from (\ref{dSvpm}) 
in terms of the superfields. We find 
\begin{eqnarray}
S_V^\pm=\mp\frac{i\k_0}{2}\int d^3z^\pm\Biggl( 
\frac{\mp\dot F^\pm+2i\Th_\mp\dot\Th_\mp}{(D_\pm\Th_\mp)^2}
{\cal D}_\pm\Th_\mp
+2i\frac{D_\pm\dot\Th_\mp D_\pm^2\Th_\mp}{(D_\pm\Th_\mp)^2}\Biggr) ~,
\label{sfSV}
\end{eqnarray}
where the superspace volume elements are defined by $d^3z^\pm=d^2xd\th_\mp$.
Under the coordinate changes (\ref{tpsct}), $S_V^\pm$ are transformed 
into $\overline S_V^\pm$ given by 
\begin{eqnarray}
\overline S_V^\pm=S_V^\pm
+2i\k_0\int d^3z^\pm\phi^\pm{\cal D}_\pm\vartheta_\mp
-4\k_0\int d^3z^\pm
\frac{D_\pm\dot\vartheta_\mp D^2_\pm\vartheta_\mp}{(D_\pm\vartheta_\mp)^2} ~. 
\label{tpSV}
\end{eqnarray}
We see that $S_V^\pm$ is invariant, up to surface terms, under the 
transformations satisfying 
\begin{eqnarray}
{\cal D}_\pm\vartheta_\mp=0 ~. \label{sMi}
\end{eqnarray}
They correspond to time-dependent super-M\"obius transformations on 
the coordinates $\s$ and $\th_\mp$. These together with the obvious 
$\tau$-reparametrization $\tau\rightarrow\overline\tau(\tau)$ constitute 
the full symmetries of the integrated super-Virasoro anomaly. 

As argued in Section 2, the string action $S_X$ at the quantum level 
is invariant under the transformations satisfying (\ref{gsl2R}), which 
are the special case of (\ref{sMi}) for infinitesimal transformations 
given by $\overline\tau=\tau+u^0(x)$, $\overline\s=\s+u^1(x) \mp2i\th_\mp
\e_\mp$ and $\overline\th_\mp=\displaystyle{\biggl(1+\frac{1}{2}
u^1{}'(x)\biggr)}\th_\mp+\e_\mp(x)$. By construction the symmetries of 
$S_X$ can be enlarged to those of (\ref{ctSV}) as mentioned before. 

\section{Super-Weyl anomaly and super-Liouville action}
\setcounter{equation}{0}

In Section 3 we have shown that the super-Virasoro anomaly 
of the original string action (\ref{str}) can be cancelled by the 
counteraction (\ref{Svpm}). The total effective action turns out to 
possess all the local symmetries of the classical theory. It is, 
however, a nonlocal functional in 2D supergravity fields. 
As was discussed in ref. \cite{fikt} for Polyakov string, we will 
introduce another counterterm which cancels the nonlocal terms of 
(\ref{Svpm}), and recover the locality of the effective action. 
We must choose the new counterterm to be invariant under 
reparametrizations and local supersymmetry to maintain these 
symmetries. 

To get some insight into the counterterm we will first investigate 
the nonlocality of (\ref{Svpm}) by using weak field expansion. Let 
us expand the 2D metric $g_{\al\be}=e_\al{}^ae_{\be a}$ by linearizing 
fields as 
\begin{equation}
g_{\al\be}=\h_{\al\be}+h_{\al\be}~. \label{wfegab}
\end{equation}
We also assume $\chi_\al$ to be the same order as $h_{\al\be}$. 
Then to lowest order in the weak fields one can easily obtain $\z_\pm$ 
and $j_\pm$ from (\ref{eqz}) and (\ref{eqj})\footnote{The light-cone 
coordinates are defined by $x^\pm
=x^0\pm x^1$. The flat metric $\h_{ab}$ in this coordinates is 
given by $\h_{++}=\h_{--}=0$, $\displaystyle{\h_{+-}=-\frac{1}{2}}$. 
We will use the convention $\partial_\pm=\partial_0\pm\partial_1$, 
hence $\partial_\pm x^\pm=2$.}
\begin{equation}
\z_\pm=\pm2\k_0\frac{\partial_\s^3}{\partial_\mp}h_{\mp\mp} ~, \qquad
j_\pm=\pm16\k_0\frac{\partial_\s^2}{\partial_\mp}\chi_{\mp\mp} ~.
\label{wfezj}
\end{equation}
Putting these into (\ref{dSvpm}) and then integrating the arbitrary 
variation of the counteraction, we find the lowest order form of $S_V$ as
\begin{eqnarray}
S_V&=&\k_0\int d^2x\Big(~\frac{1}{4}h_{++}\frac{\partial_-^3}{\partial_+}
h_{++}+\frac{1}{4}h_{--}\frac{\partial_+^3}{\partial_-}h_{--}
\nonumber\\
&&-4i\chi_{++}\frac{\partial_-^2}{\partial_+}\chi_{++}
-4i\chi_{--}\frac{\partial_+^2}{\partial_-}\chi_{--}~\Bigr)+\cdots~,
\label{wfeSv}
\end{eqnarray}
where we have supressed terms local in $h_{\al\be}$ and $\chi_\al$. 
This is an extension of the bosonic string case observed in ref. 
\cite{fikt} to fermionic string. 

As is easily inferred from the case of bosonic string, the nonlocal 
terms of (\ref{wfeSv}) can be canceled by the supersymmetric extension 
of the nonlocal Liouville action found by Polyakov \cite{pol2,tanii,ggmt,gx}. 
It can 
be defined by the super-Liouville action given by
\begin{eqnarray}
S_L&=&-\int d^2x~\Biggl[e\biggl\{~\frac{1}{2}
\bigl(g^{\al\be}\dl_\al \Phi\dl_\be \Phi
-i\overline\Psi\r^\al\nabla_\al\Psi\bigr)
+\overline\chi_\al\r^\be\r^\al\Psi\dl_\be \Phi
+\frac{1}{4}\overline\Psi\Psi\overline\chi_\al\r^\be\r^\al
\chi_\be~\biggr\} \nonumber\\
&&\hskip 2.0cm-eR\Phi
-4i\e^{\al\be}\overline\chi_\al\r_5\r^\g\chi_\g\partial_\be\Phi
-4\e^{\al\be}\overline\chi_\al\r_5\nabla_\be\Psi~\biggr]~, \label{sLact}
\end{eqnarray}
where $R$ is the scalar curvature computed from the 2D metric 
$g_{\al\be}$. The super-Liouville fields $\Phi$ and $\Psi$ are 
defined as the solutions to the classical field equations obtained 
from this action and are assumed to satisfy $\Phi=\Psi=0$ for 
$e_\al{}^a=\de_\al^a$ and $\chi_\al=0$. They depend functionally 
on the 2D supergravity fields and are, in general, nonlocal quantities. 

The lowest order approximation of (\ref{sLact}) can be computed 
in a straightforward manner. It is given by 
\begin{equation}
S_L=-\frac{1}{2}\int d^2x\Bigl\{~R\frac{1}{\Box}R
+16\overline\chi\frac{\r\cdot\partial}{\Box}\chi~\Bigr\}~,
\label{wfesL}
\end{equation}
where $R=\ddot h_{11}+h''_{00}-2\dot h_{01}'$ is the linearized 
scalar curvature and we have defined $\Box
\equiv-\partial_+\partial_-$, $\r\cdot\partial\equiv
\de^\al_a\r^a\partial_\al$ and $\chi\equiv\e^{\al\be}\partial_\al\chi_\be$. 
Then the nonlocality of (\ref{wfeSv}) can be removed within the 
linearizing approximation by introducing the new counteraction
\begin{equation}
S_T=-\frac{\k_0}{2}S_L~.\label{ST}
\end{equation}
Weak field expansion can be systematically carried out order by order 
and one can examine the cancellation of the nonlocal terms between 
$S_V$ and $S_T$ to any order, establishing the relation (\ref{ST}). 
We shall show in a moment that (\ref{ST}) is exact without relying 
on such approximation scheme.

Before going to the derivation of (\ref{ST}), we remark here the 
symmetry properties of $S_L$. It is obviously invariant under 
the reparametrization and local supersymmetry. This property is 
necessary to maintain these symmetries recovered by adding $S_V$ 
to $S_X$. It is, 
however, not invariant under the super-Weyl transformations. Under 
the infinitesimal Weyl rescaling $\displaystyle{\de e_\al{}^a
=\frac{\phi}{2}e_\al{}^a}$ and $\displaystyle{\de\chi_\al
=\frac{\phi}{4}\chi_\al}$, $S_T$ produces the Weyl anomaly, i.e.,
\begin{equation}
\de S_L=\int d^2x[~eR+4i\e^{\al\be}\partial_\al
(\overline\chi_\be\r_5\r^\g\chi_\g)~]\phi~. \label{WA}
\end{equation}
Similarly for the fermionic transformation $\de e_\al{}^a=0$ 
and $\de\chi_\al=i\r_\al\e$ with $\e$ being an infinitesimal 
Majorana field, we find the anomaly 
\begin{equation}
\de S_L=16\int d^2x~\e^{\al\be}\overline\e\r_5\nabla_\al\chi_\be~.
\label{FA}
\end{equation}
We thus see that the super-Weyl anomaly is correctly reproduced by 
requiring the locality of the counteraction.

To establish the locality of $S_V+S_T$ we will basically follow the 
idea used in ref. \cite{fikt}. We first rewite (\ref{sLact}) in 
terms of the variables (\ref{zweibein}) -- (\ref{gravitino}) as 
\begin{eqnarray}
S_L&=&\int d^2x\biggl[~\frac{(\dot \Phi-\lam^+\Phi')
(\dot \Phi+\lam^-\Phi')}{\lam^++\lam^-}
+\frac{i}{2}\Psi_+(\dot\Psi_+-\lam^+\Psi_+')
+\frac{i}{2}\Psi_-(\dot\Psi_-+\lam^-\Psi_-') \nonumber\\
&&+\frac{2}{\lam^++\lam^-}\bigl\{~i(\dot \Phi-\lam^+ \Phi')\Psi_
-\n_+-i(\dot \Phi+\lam^-\Phi')\Psi_+\n_-
+\Psi_+\Psi_-\n_+\n_- \nonumber\\
&&+\frac{1}{2}(~\dot\xi-\lam^+\xi'-2\lam^+{}'+2i\nu_-\Lam_+~)
(\dot\Phi+\lam^-\Phi') \nonumber\\
&&+\frac{1}{2}(~\dot\xi+\lam^-\xi'+2\lam^-{}'-2i\nu_+\Lam_-~)
(\dot\Phi-\lam^+\Phi'~)\bigr\} \nonumber\\
&&+i\Psi_+\bigl(\dot\Lam_+-\lam^+\Lam_+'-\frac{1}{2}\lam^+{}'\Lam_+
-4\nu_-' \nonumber\\
&&-\frac{2\nu_-}{\lam^++\lam^-}\{\dot\xi+\lam^-\xi'-(\lam^+-\lam^-)'
-i\nu_+\Lam_-~\}\bigr) \nonumber\\
&&+i\Psi_-\bigl(\dot\Lam_-+\lam^-\Lam_-'+\frac{1}{2}\lam^-{}'\Lam_-
-4\nu_+' \nonumber\\
&&-\frac{2\nu_+}{\lam^++\lam^-}\{\dot\xi-\lam^+\xi'-(\lam^+-\lam^-)'
-i\nu_-\Lam_+~\}\bigr)~\biggr]~, \nonumber\\
\label{sLact2}
\end{eqnarray}
and then perform the supertranslation by $\h_\mp$ satisfying 
(\ref{fpm}), hence $\hat\nu_\mp=0$. In this special gauge 
the bosonic and fermionic sectors are decoupled in (\ref{sLact2}) and 
one can easily observe that 
\begin{equation}
\hat\Lam\equiv\pmatrix{(-\hat e_1{}^-)^{-1/2}\hat\Lam_- \cr
(\hat e_1{}^+)^{-1/2}\hat\Lam_+ \cr}
\label{hatLam}
\end{equation}
transforms as a scalar under reparametrizations and as a spinor 
under local Lorentz transformaitons. The action can be simplified 
further if we work in the new coordintates $(\tilde x)$ defined by
\begin{equation}
\tilde x^\pm=f_\mp(x) ~,\label{hcoord}
\end{equation}
where $f_\pm(x)$ are introduced in Section 3. As in the 
bosonic string case, the metric tensor becomes conformally minkowskian 
and is given by
\begin{equation}
\tilde g_{\al\be}(\tilde x)
=\h_{\al\be}\frac{\hat g_{11}(x)}{f'_+(x)f'_-(x)}~. \label{cmmetr}
\end{equation}
The super-Liouville action (\ref{sLact2}) reduces to 
\begin{eqnarray}
S_L&=&\int d^2\tilde x\bigl\{~-\frac{1}{2}
\h_{\al\be}\tilde\partial_\al\tilde\Phi\tilde\partial_\be\tilde\Phi
+\frac{i}{2}\tilde\Psi_+(\dot{\tilde\Psi}_+-\tilde\Psi'_+)
+\frac{i}{2}\tilde\Psi_-(\dot{\tilde\Psi}_-+\tilde\Psi'_-) \nonumber\\
&&-\h_{\al\be}\tilde\partial_\al\tilde\partial_\be\tilde\xi\tilde\Phi
-i\tilde\Lam_+(\dot{\tilde\Psi}_+-\tilde\Psi'_+)
-i\tilde\Lam_-(\dot{\tilde\Psi}_-+\tilde\Psi'_-)\bigr\}~. \label{sLact3}
\end{eqnarray}
This immediately gives the classical solution for the super-Liouville 
fields as
\begin{equation}
\tilde\Phi(\tilde x)=\tilde\xi(\tilde x), \qquad
\tilde \Psi_\pm(\tilde x)=\tilde\Lam_\pm(\tilde x)~. \label{tcsol}
\end{equation}
Since the super-Liouville fields $\Phi$, $\Psi$ and $\hat\Lam$ defined by 
(\ref{hatLam}) transform as scalars under world-sheet reparametrizations, 
we easily obtain from (\ref{tcsol}) the super-Liouville fields in the 
original coordinates $(x)$ as
\begin{equation}
\hat\Phi(x)=\hat\xi(x)-\ln f'_+(x)f'_-(x), \qquad
\hat\Psi_\pm(x)=\hat\Lam_\pm(x)~. \label{hcsol}
\end{equation}
In deriving these we have used the relations (\ref{cmmetr}). The 
super-Liouville action $S_L$ can be obtained 
from (\ref{sLact3}) by substituting the classical solutions (\ref{tcsol}) 
or (\ref{hcsol}) for the super-Liouville fields. In the $(x)$-coordinates 
all the nonlocalities of $S_L$ arise through the dependences of various 
variables on $f_\pm$ and $\h_\mp$. We would like to separate the dependences
on these variables. 

To do this we first note that the scalar curvature density in the 
$(\tilde x)$-coordinates is given by 
$\tilde e\tilde R=-\h_{\al\be}\partial_\al\partial_\be\tilde\xi$. 
Putting this into the rhs of (\ref{sLact3}) and coming back to the 
$(x)$-coordinate system, we obtain 
\begin{equation}
S_L=S_{\rm a}+S_{\rm b}~, \label{sLact4}
\end{equation}
where $S_{\rm a,b}$ are given by 
\begin{eqnarray}
S_{\rm a}&=&\int d^2x\Big\{~\hat e\Bigl(~-\frac{1}{2}\hat g^{\al\be}
\partial_\al\hat\xi\partial_\be\hat\xi+\hat R\hat\xi~\Bigr)
-\frac{(\hat\lam^+{}'-\hat\lam^-{}')^2}{\hat\lam^++\hat\lam^-} \nonumber\\
&&-\frac{i}{2}\hat\Lam_+(\dot{\hat\Lam}_+-\hat\Lam_+')
-\frac{i}{2}\hat\Lam_-(\dot{\hat\Lam}_-+\hat\Lam_-')~\Bigr\}~, \label{Sa}\\
S_{\rm b}&=&\int d^2x\Bigl(~
\frac{(f_+'')^2\dot f_+}{(f_+')^3}-\frac{\dot f_+'f_+''}{(f_+')^2}
-\frac{(f_-'')^2\dot f_-}{(f_-')^3}+\frac{\dot f_-'f_-''}{(f_-')^2}
~\Bigl)~. \label{Sb}
\end{eqnarray}
We see that except for the second term in the integrand of (\ref{Sa}) 
$S_{\rm a}$ can be obtained from the super-Liouville action 
(\ref{sLact2}) by the following replacements
\begin{eqnarray}
&&\lam^\pm,~\xi,~\nu_\mp,~\Lam_\pm\rightarrow
\hat\lam^\pm,~\hat\xi,~\hat\nu_\mp(=0),~\hat\Lam_\pm, \nonumber\\
&&\Phi,~\Psi\rightarrow\hat\xi,~\hat\Lam_\pm~. \label{rpl}
\end{eqnarray}
Hence, let us define a family of variables parametrized by 
a real parameter $\al$ by the set of differential equations
\begin{eqnarray}
&&\frac{d\lam^\pm(\al)}{d\al}=-4i\h_\mp\nu_\mp(\al)~, \nonumber\\
&&\frac{d\nu_\mp(\al)}{d\al}=\dot\h_\mp\mp\lam^\pm(\al)\h_\mp'
\pm\frac{1}{2}\lam^\pm{}'(\al)\h_\mp~, \nonumber\\
&&\frac{d\xi(\al)}{d\al}=i(\h_+\Lam_-(\al)-\h_-\Lam_+(\al))~, \nonumber\\
&&\frac{\lam^+(\al)+\lam^-(\al)}{4}\frac{d\Lam_\pm(\al)}{d\al}=
(\lam^+(\al)+\lam^-(\al))\h_\mp'
\mp\frac{1}{2}(\lam^+(\al)-\lam^-(\al))'\h_\mp \nonumber\\
&&\hskip 2.5cm\pm\frac{1}{2}\h_\mp(\dot\xi(\al)\pm\lam^\pm\xi')
\pm\frac{i}{2}(\Lam_-(\al)\nu_+(\al)-\Lam_+(\al)\nu_-(\al))\h_\mp~, 
\nonumber\\
&&\frac{d\Phi(\al)}{d\al}=i(\h_+\Psi_-(\al)-\h_-\Psi_+(\al))~, \nonumber\\
&&\frac{\lam^+(\al)+\lam^-(\al)}{4}\frac{d\Psi_\pm(\al)}{d\al}=
(\lam^+(\al)+\lam^-(\al))\h_\mp'
\mp\frac{1}{2}(\lam^+(\al)-\lam^-(\al))'\h_\mp \nonumber\\
&&\hskip 2.5cm\pm\frac{1}{2}\h_\mp(\dot\Phi(\al)\pm\lam^\pm\Phi')
\pm\frac{i}{2}(\Psi_-(\al)\nu_+(\al)-\Psi_+(\al)\nu_-)\h_\mp~, 
\nonumber\\
&&\label{1pfam}
\end{eqnarray}
with the initial conditions
\begin{equation}
\lam^\pm(0)=\lam^\pm~, \quad \nu_\mp(0)=\nu_\mp~, \quad
\xi(0)=\Phi(0)=\xi~,\quad \Lam_\pm(0)=\Psi_\pm(0)=\Lam_\pm~. \label{inicon}
\end{equation}
Except for the equations for $\Phi(\al)$ and $\Psi_\pm(\al)$,
(\ref{1pfam}) define a one-parameter family of 2D metrices and gravitinos
connected by a local supersymmetry. At $\al=1$, we obtain 
$\lam^\pm(1)=\hat\lam^\pm$, $\xi(1)=\hat\xi$ and $\nu_\mp(1)=\hat\nu_\mp=0$.
On the other hand the last two equations together with the initial 
conditions (\ref{inicon}) give
\begin{equation}
\Phi(\al)=\xi(\al)~, \qquad \Psi_\pm(\al)=\Lam_\pm(\al)~. \label{PPcons}
\end{equation}
Explicit solutions to (\ref{1pfam}) can be found in a straightforward 
manner and turn out to be polynomials (at most cubic) in $\al$. For 
instance, $\nu_\mp(\al)$ are given by
\begin{equation}
\nu_\mp(\al)=\nu_\mp+\al(\dot\h_\mp\mp\lam^\pm\h_\mp'
\pm\frac{1}{2}\lam^\pm{}'\h_\mp)\mp3i\al^2\h_\mp\h_\mp'\nu_\mp
\mp i\al^3\h_\mp\h_\mp'\dot\h_\mp~. \label{nual}
\end{equation}

Let us now define a one-parameter family of actions by 
\begin{equation}
S(\al)=S_L(\al)-\int d^2x~
\frac{(\lam^+{}'(\al)-\lam^-{}'(\al))^2}{\lam^+(\al)+\lam^-(\al)}~,
\label{Sal}
\end{equation}
where $S_L(\al)$ is obtained form (\ref{sLact2}) by replacing 
all the variables by the corresponding $\al$-dependent ones introduced 
above. With this definition, we find 
\begin{equation}
S_{\rm a}=S(1)~. \label{Sa2}
\end{equation}

To find the explict form of (\ref{Sal}) we first derive the differential 
equation satifsfied by $S(\al)$. This can be found by the observation 
that $S_L(\al)$ would be independent of $\al$ if $\Psi_\pm(\al)$ obeyed 
the equation
\begin{equation}
\frac{\lam^+(\al)+\lam^-(\al)}{4}\frac{d\Psi_\pm(\al)}{d\al}=
\pm\frac{1}{2}\h_\mp(\dot\Phi(\al)\pm\lam^\pm\Phi')
\pm\frac{i}{2}(\Psi_-(\al)\nu_+(\al)-\Psi_+(\al)\nu_-)\h_\mp~. \label{wdbe}
\end{equation}
This implies that $S(\al)$ is subject to the differential equation
\begin{equation}
\frac{dS(\al)}{d\al}=\int d^2x\Bigl[~
\Delta\Psi_+(\al)\frac{\de S(\al)}{\de\Psi_+(\al)}
+\Delta\Psi_-(\al)\frac{\de S(\al)}{\de\Psi_-(\al)}
-\frac{d}{d\al}
\frac{(\lam^+{}'(\al)-\lam^-{}'(\al))^2}{\lam^+(\al)+\lam^-(\al)}
~\Bigr]~, \label{dSal}
\end{equation}
where $\Delta\Psi_\pm(\al)$ stand for the deviations of the last 
equations in (\ref{1pfam}) from (\ref{wdbe}), i.e.,
\begin{equation}
\Delta\Psi_\pm(\al)\equiv 4\h_\mp'\mp
\frac{2(\lam^+(\al)-\lam^-(\al))'}{\lam^+(\al)+\lam^-(\al)}\h_\mp~.
\label{exterm}
\end{equation}
Surprisingly enough, many terms are canceled out in the rhs of 
(\ref{dSal}) due to the supersymmetry mentioned above and we find 
\begin{equation}
\frac{dS(\al)}{d\al}=16i\int d^2x(~\nu_-(\al)\h_-''+\nu_+(\al)\h_+''~)~.
\label{dSal2}
\end{equation}
This just corresponds to the anomaly equation (\ref{qcovconsJ}) for the 
local supersymmetry. 

Using (\ref{nual}), we can integrate (\ref{dSal2}) to obtain 
\begin{eqnarray}
S_{\rm a}&=&S_L^\xi-\int d^2x\frac{(\lam^+{}'-\lam^-{}')^2}{\lam^++\lam^-} 
\nonumber\\
&&-4\int d^2x\Bigl[~2i\dot\h_-\h_-''
-i\frac{\dot f_+}{f_+'}(3\h_-'\h_-''+\h_-\h_-''')
-\h_-\h_-'\h_-''\dot\h_- \nonumber\\
&&+2i\dot\h_+\h_+''
-i\frac{\dot f_-}{f_-'}(3\h_+'\h_+''+\h_+\h_+''')
+\h_+\h_+'\h_+''\dot\h_+~\Bigr]~. \label{Sich}
\end{eqnarray}
In deriving (\ref{Sich}), we have used the relations (\ref{fpm}).
The $S_L^\xi$ is defined from (\ref{sLact2}) with the replacements 
$\Phi\rightarrow\xi$ and $\Psi_\pm\rightarrow\Lam_\pm$, hence is a 
local functional of zweibein and gravitino. The nonlocalities 
of $S_{\rm a}$ are completely isolated as terms containing $f_\pm$ 
and $\h_\mp$. Combining (\ref{ctSV}), (\ref{Svpm}), (\ref{ST}), 
(\ref{sLact4}), (\ref{Sich}) and (\ref{Sb}), we finally obtain
\begin{eqnarray}
S_g&\equiv& S_V+S_T \nonumber\\
&=&\frac{\k_0}{2}\int d^2x\Biggl[~e\Bigl\{~\frac{1}{2}
(g^{\al\be}\partial_\al\xi
\partial_\be\xi-i\overline\Lam\r^\al\nabla_\al\Lam)+
\overline\chi_\al\r^\be\r^\al\Lam\partial_\be\xi
+\frac{1}{4}\overline\Lam\Lam\chi_\al\r^\be\r^\al\chi_\be~\Bigr\} 
\nonumber\\
&&-eR\xi-4i\e^{\al\be}\overline\chi_\al\r_5\r^\g\chi_\g\partial_\be\xi
-4\e^{\al\be}\overline\chi_\al\r_5\nabla_\be\Lam
+\frac{2g_{11}}{\sqrt{-g}}
\Bigl\{\Bigl(\frac{g_{01}}{g_{11}}\Bigr)'\Bigr\}^2~\Biggr]~, \label{Sg}
\end{eqnarray}
where $\Lam$ is defined by 
\begin{equation}
\Lam\equiv \pmatrix{(-e_1{}^-)^{-1/2}\Lam_- \cr
(e_1{}^+)^{-1/2}\Lam_+\cr}~. \label{Lam}
\end{equation}
This is the local super-Liouville action with $\xi$, $\Lam$ identified 
with the super-Liouville fields and has been obtained in 
\cite{fikkt,fir-kt}. 
As discussed in \cite{fikkt,fir-kt}, $S_g$ is not invariant under 
world-sheet 
reparametrizations and 
local supersymmetry since the super-Liouville fields $\xi$, $\Lam$ do not 
possess simple transformation properties such as scalar and spinor. 
Instead it exactly cancels the super-Virasoro anomaly given in 
(\ref{qcovconsvp}) and (\ref{qcovconsJ}). Furthermore, we see from 
(\ref{WA}) and (\ref{FA}) that the super-Weyl anomaly is correctly 
reproduced by (\ref{Sg}). In this sense $S_g$ can be regarded as a 
WZW term converting the super-Virasoro anomaly into 
the super-Weyl anomaly. It can also be considered as a seagull term 
for the action (\ref{str}) to maintain the covariance. The requirement 
of locality, which is dictated by the perturbative analysis, is so 
restrictive that we can not maintain all the local symmetries of the 
classical action. We stress that the action (\ref{Sg}) is local in 
contrast to the nonlocal action (\ref{sLact}). This is only 
possible by sacrificing the covariance under reparametrizations and 
supersymmetry. 

We will close this section by mentioning the inclusion of the 
auxiliary fields for the supermultiplets and the cosmological 
term. The latter arises through quantum effects, giving additional 
contributions to the super-Weyl anomaly. The auxiliary fields should 
be introduced to retain off-shell closure of the supersymmetry algebra, 
which ensures the off-shell nilpotency of the BRST transformations. 
As was stressed in ref. \cite{fir-kt}, this is crucial in order to acheive 
quantization of the model for general gauge-fixings. The action 
$S_X+S_g$, however, is not invariant under the supersymmetry 
transformations modified to satisfy off-shell closure condition 
becuase of the super-Weyl anomaly. This can be remedied by 
introducing additional term to the action which is quadratic in 
the auxiliary field for the zweibein multiplet and turns out to 
be related with the cosmological term \cite{pol2,ft,fir-kt}. Thus 
the additional action 
including the cosmological term is given by
\begin{eqnarray}
S_{\rm cosm}&=&\frac{\k_0}{2}\int d^2xe\Bigl[\frac{1}{2}A^2+2\m(A-2e^{-1}
\e^{\al\be}\overline\chi_\al\r_5\chi_\be)\Bigr] \nonumber\\
&=&\frac{\k_0}{2}\int d^2x\Bigl[\frac{1}{2}a^2
+\m e^\frac{\xi}{2}\sqrt{2(\lam^++\lam^-)}a
-2i\m e^\frac{\xi}{2}(\n_+\Lam_++\n_-\Lam_-) \nonumber\\
&&+\frac{i\m}{2}e^\frac{\xi}{2}
(\lam^++\lam^-)\Lam_-\Lam_+\Bigr]~, \label{cosm1}
\end{eqnarray}
where $A$ is the auxiliary field for the zweibein supermultiplet. We 
have introduced the rescaled variable $a$ given by
\begin{eqnarray}
a=\sqrt{e}A ~ \label{a} 
\end{eqnarray}
for later convenience. The auxiliary field can be eliminated by using 
the equation of motion $A=-2\m$, leading 
to the standard form for the cosmological term
\begin{eqnarray}
S_{\rm cosm}=-\k_0\int d^2x(\m^2e
+2\m\e^{\al\be}\overline\chi_\al\r_5\chi_\be)~. \label{cosm2}
\end{eqnarray}
We should also include auxiliary fields for the string sector. Since 
they play no role in the subsequent section, we omit them here. 

The action (\ref{Sg}) toghether with (\ref{cosm1}) describe the dynamics 
of the super-Liouville mode of 2D supergravity. We may then regard 
$S_X+S_g+S_{\rm cosm}$ as starting action to canonically quantize 
this system, which we will argue in the next section.

\section{BRST Quantization}
\setcounter{equation}{0}

So far we have considered the zweibeins and the gravitinos as background 
classical fields and only the string variables are treated as quantized 
operators. In the previous sections we have shown that the quantum 
fluctuations of the string variables induce the super-Liouville action 
(\ref{Sg}), which can be considered to describe the dynamics of the 
super-Liouville mode of the 2D supergravity. In this section we will 
promote them to dynamical variables and investigate their quantization. 
Since graviton and gravitino themselves including the superghosts 
associated with gauge-fixing also contribute to the super-Virasoro 
anomaly, we will replace the coupling constant $\k_0$ by an arbitrary 
parameter $\k$ in the starting action 
\begin{eqnarray}
S_0=S_X+S_g+S_{\rm cosm} ~. \label{S0}
\end{eqnarray}
Then, despite the apparent lack of invariances under reparametrizations 
and local supersymmetry, they will recover at the quantum level for 
an appropriate choice of $\k$. We will argue the BRST invariance of the 
theory described by the action in the superconformal gauge and in 
the supersymmetric light-cone gauge. Since these gauges have already 
been taken up in ref. \cite{fir-kt} for essentially the same action, 
we will be 
brief and contain them here to make the paper self-contained. 

\subsection{Superconformal gauge-fixing}

The Superconformal gauge is defined by the relations 
\begin{eqnarray}
e_\al{}^a=\sqrt{e}\de_\al^a~, \qquad 
\chi_\al=-\frac{1}{2}\r_\al\r^\be\chi_\be ~. \label{scg2}
\end{eqnarray}
These correspond to the gauge conditions (\ref{scg}). We summarize 
the BRST transformations in covariant notation in Appendix A. The 
BRST gauge-fixed action in this gauge is then given by
\begin{eqnarray}
S_{\rm eff}&=&\int d^2x\Bigl[\frac{1}{2}\dl_+X\dl_-X
+\frac{i}{2}\psi_+\dl_-\psi_++\frac{i}{2}\psi_-\dl_+\psi_-\Bigr] \nn \\
&& +\frac{\k}{2}\int d^2x\Bigl[\frac{1}{2}\dl_+\xi\dl_-\xi 
+\frac{i}{2}\Lam_+\dl_-\Lam_++\frac{i}{2}\Lam_-\dl_+\Lam_-
+\m e^{\frac{\xi}{2}}(2a+i\Lam_-\Lam_+)+\frac{1}{2}a^2\Bigr] \nn \\
&& +\int d^2x [ -b_{++}\dl_-c^+-b_{--}\dl_+c^-
+\be_{++}\dl_-\g_-+\be_{--}\dl_+\g_+] ~, \label{gfascg}
\end{eqnarray}
where $c^\pm\equiv C^\pm$ and $\g_\mp\equiv\w_\mp$ are superconformal 
ghosts, and $b_{\pm\pm}$ 
and $\be_{\pm\pm}$ are their antighosts. Since the local Lorentz mode 
$l$ and the associated ghost fields are not propagating degrees, they 
have been integrated out via the equations of motion from the action. 

The action (\ref{gfascg}) is invariant under the BRST transformations 
\begin{eqnarray}
\de X&=& \frac{1}{2}c^+\dl_+X+\frac{1}{2}c^-\dl_-X
-i(\g_-\psi_+-\g_+\psi_-) ~, \nn \\
\de \psi_\pm&=&\frac{1}{2}c^\pm\dl_\pm\psi_\pm
+\frac{1}{4}\dl_\pm c^\pm\psi_\pm\pm\g_\mp\dl_\pm X ~, \nn \\
\de\xi&=&\frac{1}{2}\dl_+c^++\frac{1}{2}\dl_-c^-
+\frac{1}{2}c^+\dl_+\xi+\frac{1}{2}c^-\dl_-\xi
-i(\g_-\Lam_+-\g_+\Lam_-) ~, \nn \\
\de \Lam_\pm&=&\frac{1}{2}c^\pm\dl_\pm\Lam_\pm
+\frac{1}{4}\dl_\pm c^\pm\Lam_\pm\pm\g_\mp\dl_\pm\xi
\pm2\dl_\pm\g_\mp\pm\frac{\m}{2}e^{\frac{\xi}{2}}c^\mp\Lam_\mp
-2\m e^{\frac{\xi}{2}}\g_\pm ~, \nn \\
\de a&=&\frac{1}{2}c^+\dl_+a+\frac{1}{2}c^-\dl_-a
+\frac{1}{4}\dl_+c^+a+\frac{1}{4}\dl_-c^-a
-i(\g_+\dl_-\Lam_++\g_-\dl_+\Lam_-) ~, \nn \\
\de c^\pm&=&\frac{1}{2}c^\pm\dl_\pm c^\pm+2i\g_\mp^2 ~, \nn \\
\de\g_\pm&=&\frac{1}{2}c^\mp\dl_\mp \g_\pm
-\frac{1}{4}\dl_\mp c^\mp\g_\pm ~, \nn \\
\de b_{\pm\pm}&=&T^X_{\pm\pm}+T^L_{\pm\pm}+T^{gh(2)}_{\pm\pm}
+T^{gh(3/2)}_{\pm\pm} ~, \nn \\
\de \be_{\pm\pm}&=&\pm i(J^X_{\pm\pm}+J^L_{\pm\pm}+J^{gh}_{\pm\pm}) ~,
\label{scbrsttr}
\end{eqnarray}
where the stress tensors $T^{X,L,gh(2,3/2)}_{\pm\pm}$ and the supercurrents 
$J^{X,L,gh}_{\pm\pm}$ are given by
\begin{eqnarray}
T^X_{\pm\pm}&\equiv&\frac{1}{4}(\dl_\pm X)^2
+\frac{i}{4}\psi_\pm\dl_\pm\psi_\pm ~, \nn \\
T^L_{\pm\pm}&\equiv&\frac{\k}{2}\Bigl[\frac{1}{4}(\dl_\pm\xi)^2
-\frac{1}{2}\dl_\pm^2\xi+\frac{i}{4}\Lam_\pm\dl_\pm\Lam_\pm\Bigr] ~, \nn \\
T^{gh(2)}_{\pm\pm}&\equiv&-b_{\pm\pm}\dl_\pm c^\pm
-\frac{1}{2}\dl_\pm b_{\pm\pm}c^\pm ~, \nn \\
T^{gh(3/2)}_{\pm\pm}&\equiv&\frac{3}{4}\be_{\pm\pm}\dl_\pm\g_\mp
+\frac{1}{4}\dl_\pm\be_{\pm\pm}\g_\mp ~, \nn \\
J^X_{\pm\pm}&\equiv&\psi_\pm\dl_\pm X ~, \nn \\
J^L_{\pm\pm}&\equiv&\frac{\k}{2}(\Lam_\pm\dl_\pm\xi-2\dl_\pm\Lam_\pm) ~, 
\nn \\
J^{gh}_{\pm\pm}&\equiv&\mp i\Bigl(\frac{3}{4}\be_{\pm\pm}\dl_\pm c^\pm
+\frac{1}{2}\dl_\pm\be_{\pm\pm}c^\pm-4ib_{\pm\pm}\g_\mp\Bigr) ~.
\label{stsc}
\end{eqnarray}
Careful reader might be sceptical over the BRST invariance of 
(\ref{gfascg}) since the action $S_g$ is neither 
generally covariant nor locally supersymmetric at the classical 
level. This can be understood by noticing that the super-Virasoro 
anomaly appreaing in (\ref{eqz}) and (\ref{eqj}) vanishes in this 
gauge as pointed out in Section 2. 

The BRST charge generating the transformations (\ref{scbrsttr}) is 
given by 
\begin{eqnarray}
Q_B&=&\int d\s \Bigl[c^+(T^X_{++}+T^L_{++}+\frac{1}{2}T^{gh(2)}_{++}
+T^{gh(3/2)}_{++})+c^-(T^X_{--}+T^L_{--}+\frac{1}{2}T^{gh(2)}_{--}
+T^{gh(3/2)}_{--}) \nn \\
&& -i\g_-(J_{++}^X+J_{++}^L)
+i\g_+(J_{--}^X+J_{--}^L)+2ib_{++}\g_-^2+2ib_{--}\g_+^2\Bigr] ~, 
\label{scQB}
\end{eqnarray}
which is conserved correspondingly to the invariance of the action but 
not nilpotent under super-Poisson bracket, {\it i.e.,} $\{Q_B,Q_B\}
\neq0$, even at the classical level due to the nonvanishing central 
term in the super-Virasoro algebra for the super-Liouville sector. 

Quantum theory can be acheived by replacing fundamental super-Poisson 
brackets $\{~,~\}$ with supercommutator $\displaystyle{\frac{1}{i}[~,~]}$. 
To define operator products, free field normal ordering can be used 
for string coordinates 
and ghost fields. As for the super-Liouville sector, we should be careful 
in defining $T^L_{\pm\pm}$ and $J^L_{\pm\pm}$. Since we wish to retain 
superconformal symmetry, they must be defined to satisfy super-Virasoro 
algebra. One way to implement this is to reduce the super-Liouville fields 
to free ones by choosing the cosmological term to zero. In this case the 
stress tensor and the supercurrent become those of free field with coupling 
to background charge as in refs. \cite{dk,dhk}. 

Another way to define these operators is to apply ordering prescription 
based on the decomposition (\ref{creanniop}) 
to the super-Liouville fields. We describe in some detail the derivation 
of super-Virasoro algebra in Appendix B. We show there that the Liouville 
sector contribute 
to the additional Virasoro central charge by $3/2$ due to the quantum 
fluctuation and gravitational dressing effect \cite{d,dk}, which was 
originally observed in ref. \cite{ct} for Liouville theory, is argued. 
Then the 
BRST charge satisfies the nilpotency for the vanishing total 
central charge. This leads to the condition for the coupling $\k$
\begin{eqnarray}
\k=\frac{9-D}{16\p} ~.\label{kappa}
\end{eqnarray}
As is shown in Appendix B, the cosmological term in the super-Liouville 
action suffers from gravitational dressing effect to retain BRST 
invariance. We then arrive at the super-Liouville action
\begin{eqnarray}
S_g+S_{\rm cosm}&=&\frac{\k}{2}\int d^2x\Bigl[\frac{1}{2}\dl_+\xi\dl_-\xi 
+\frac{i}{2}\Lam_+\dl_-\Lam_++\frac{i}{2}\Lam_-\dl_+\Lam_- \nn \\
&&+2\m e^{\al\xi}(a+i\al\Lam_-\Lam_+)+\frac{1}{2}a^2\Bigr] \nn \\
&=&\frac{\k}{2}\int d^2xd^2\th\Bigl( iD_+\Xi D_-\Xi+
\frac{4\m}{\al}e^{\al\Xi}\Bigr)~,
\label{qSL}
\end{eqnarray}
where we have defined a superfield\footnote{The $\th_\mp$ and 
$\displaystyle{D_\mp=\mp i\frac{\dl}{\dl\th_\pm}\mp \th_\pm\dl_\mp}$, 
respectively, stand for the fermionic coordinates and covariant 
derivatives of the flat superspace.} $\Xi(x,\th)=\xi 
+i(\th_+\Lam_--\th_-\Lam_+)+i\th_+\th_-a$. The parameter $\al$, 
which is classically $\frac{1}{2}$, is determined by the condition 
that the operator $e^{\al\Xi}$ transforms as 
superconformal field of weight $(\frac{1}{2},\frac{1}{2})$ so that 
the integral over the superspace becomes invariant under superconformal 
transformations. This combined with (\ref{kappa}) 
leads to 
\begin{eqnarray}
\al=\frac{9-D\pm\sqrt{(1-D)(9-D)}}{4} ~. \label{alpha}
\end{eqnarray}
By the rescaling $\displaystyle{\sqrt{\frac{9-D}{8}}\Xi \rightarrow \Xi}$, 
we get the standard normarization for the action (\ref{qSL}). We thus 
arrive at the minkowskian version of the principal results of refs. 
\cite{dk,dhk}. The well-known $c=1$ barrier is also observed through 
the definition of the cosmological term. In other respects our arguments 
seem to be insensitive in going beyond the barrier. 

\subsection{Supersymmetric light-cone gauge}
Light-cone gauge is defined by the conditions
\begin{eqnarray}
e_+{}^+=e_-{}^-=1~,\quad e_-{}^+=0~, \quad 
\chi_{-+}=\chi_{--}=0 ~. \label{lcgc}
\end{eqnarray}
In this gauge our parametrizations (\ref{zweibein}), (\ref{gravitino}) 
and (\ref{a}) are reduced to 
\begin{eqnarray}
&&\lam^+=1~, \qquad\qquad\quad\; \n_-=0~, \;\qquad\qquad\qquad
\Lam_+=4\chi_{+-}~, \nn\\
&&\lam^-=\frac{1-g_{++}}{1+g_{++}}~, \qquad 
\n_+=\frac{2\chi_{++}}{(1+g_{++})^\frac{3}{2}}~, \qquad
\Lam_-=\frac{4\chi_{++}}{\sqrt{1+g_{++}}}~, \nn \\
&& \xi={\rm ln}(1+g_{++})~, ~\quad 
l=-\frac{1}{2}{\rm ln}(1+g_{++}) ~, \;\quad a=A ~, \label{plcg}
\end{eqnarray}
where $e_+{}^-=-g_{++}$ and $\chi_{+\pm}$ are the unfixed gravitational 
dynamical degrees. 

To write down the gauge-fixed action is again a standard 
routine. In ref. \cite{fir-kt} a slightly different gauge conditions, 
which are not 
manifestly invariant under the rigid Lorentz rotations on the two 
dimensional 
parameter space, are employed. Here we will adopt the Lorentz invariant 
guage conditions (\ref{lcgc}) to make it manifest the rigid Lorentz 
covariance 
of the ghost and gauge-fixing parts of the action. We describe the 
handling of the ghost sector in Appendix C in some detail. We also show 
that all the ghost variables can be made free fields by suitable 
field redefinitions. As a sequel, the gauge-fixed action is found to be 
\begin{eqnarray}
S_{\rm eff}&=&\int d^2x \Bigl[~\frac{1}{2}\dl_-X(\dl_+X+g_{++}\dl_-X)
+\frac{i}{2}\psi_+\dl_-\psi_+ \nn \\
&&+\frac{i}{2}\psi_-(\dl_+\psi_-+g_{++}\dl_-\psi_-) 
-2i\chi_{++}\psi_-\dl_-X\Bigr] \nn \\
&& +\frac{\k}{2}\int d^2x \Bigl[~\frac{1}{2(1+g_{++})}\{(\dl_-g_{++})^2
-2\dl_-g_{++}
(\ln(1+g_{++}))'+4(\ln(1+g_{++}))''\} \nn \\
&&+\frac{8i}{1+g_{++}}\chi_{++}\Biggl(\dl_-\chi_{++}
-\frac{2\chi_{++}'}{1+g_{++}}\Biggr)
+\frac{i}{2}\chi_+\dl_-\chi_+\Bigr] \nn \\
&& +\int d^2x(-b_{++}\dl_-c^+-b\dl_-c_++\be_{++}\dl_-\g_-
+\be_{+}\dl_-\g_+)~, \label{lcgfseff}
\end{eqnarray}
where the $(-e_1{}^-)^\frac{1}{2}\psi_-$ given in (\ref{fermion}) has 
been redefined as $\psi_-$ and all the nonpropagating fields have been 
integrated out via the equations of motion. We have also set the 
cosmological mass $\m$ to zero for simplicity.\footnote{Though the 
cosmological term reduces to a constant in this gauge, it contributes 
to the equations of motion, hence to the BRST charge.} 

The gravitational part of the effective action can be endowed 
with an interpretation only in terms of $f_-(x)$ and $\h_+(x)$ 
introduced in Section 3. As can be seen from (\ref{fpm}) and 
(\ref{plcg}), the $f_+$ and $\h_-$ come to be decoupled from the 
gravitational variables in this gauge, and hence can be ignored 
from the beginning. The super-Liouville 
equations for $\Phi$ and $\Psi_\mp$ described by (\ref{sLact}) can 
be solved in terms of $f_-$ and $\h_+$ as
\begin{eqnarray}
\Phi=-\ln \dl_-f_-+i\xi_+\dl_-\xi_+ ~, \quad
\Psi_-=-\frac{2}{\sqrt{\dl_-f_-}}\dl_-(\sqrt{\dl_-f_-}\xi_+) ~, \qquad
\Psi_+=4\chi_{+-} ~, \label{cLeqsol}
\end{eqnarray}
with $\xi_+\equiv\sqrt{-e_1{}^-}\h_+$. Inserting these solutions into 
(\ref{sLact}), we obtain supersymmetric extension of the gravitational 
WZW action for the bosonic string case given in ref. \cite{pol1}. 
It is interesting 
to note that the resulting super-Liouville action coincides, up to over 
all constant and $\chi_{+-}$ kinetic term, with the one obtained from 
$S_V^+$ by the replacements $\dl_\tau,~\dl_\s,~f_+~, \h_-\rightarrow 
\frac{1}{2}\dl_+,~\frac{1}{2}\dl_-,~f_-,~\xi_+$. To see this explicitly, 
let us introduce light-cone superspace with supercoordinates 
$z=(x^+,x^-,\th_+)$ 
and supercovariant derivative $\displaystyle{D_-=-i\frac{\dl}{\dl\th_+}
-\th_+\dl_-}$ \cite{bmg}. The graviton superfield is defined by 
${\cal G}_{++}(z^+)=
g_{++}(x)-4i\th_+\chi_{++}(x)$. In terms of the superfields 
${\cal F}_-(z)$ and 
$\Xi_-(z)$ given by
\begin{eqnarray}
{\cal F}_-(z)&=&f_-(x)-i\th_+\xi_+(x)\dl_-f_-(x) ~, \nn\\
\Xi_-(z)&=&\sqrt{\dl_-f_-(x)}\{\xi_+(x)+\th_+(1-\frac{i}{2}
\xi_+(x)\dl_-\xi_+(x))\} ~,
\label{lcscsf}
\end{eqnarray}
the ${\cal G}_{++}$ can be written as 
\begin{eqnarray}
{\cal G}_{++}=\frac{\dl_+{\cal F}_--i\Xi_-\dl_+\Xi_-}{(D_-\Xi_-)^2} ~.
\end{eqnarray}
The super-Liouville action (\ref{sLact}) is then given by 
\begin{eqnarray}
-\frac{\k}{2}S_L&=&i\k\int d^2xd\th_+\Biggl(
\frac{\dl_+{\cal F}_--i\Xi_-\dl_+\Xi_-}{(D_-\Xi_-)^2}{\cal D}_-\Xi_-
-i\frac{\dl_+D_-\Xi_-D_-^2\Xi_-}{(D_-\Xi_-)^2}\Biggr) +\cdots ~, \nn\\ 
\label{nlact}
\end{eqnarray}
where we have suppressed the $\chi_{+-}$ kinetic term. This result 
has already been noted in ref. \cite{gx,gn}. 
The general 
argument presented in the previous section shows that the nonlocalities 
of (\ref{nlact}) cancel those of $S_V^-$ and the local 
terms left over by the cancellation constitute the gravitational action 
in (\ref{lcgfseff}). 
The light-cone gauge is retained under the pseudo-superconformal 
transformations $z=(x^+,~x^-,~\th_+) \rightarrow z'=(x^+,~g_+(z),
~\vartheta_+(z))$ satisfying $D_-g_+=-2i \vartheta_+D_-\vartheta_+$ 
if the ${\cal G}_{++}$ is subject to the relation
\begin{eqnarray}
{\cal G}'_{++}(z')=(iD_-\vartheta_+(z))^2{\cal G}_{++}(z)+\frac{1}{2}
\{\dl_+g_+(z)-2i\vartheta_+(z)\dl_+\vartheta_+(z)\} ~. \label{lcgppct}
\end{eqnarray}
The super-Liouville action (\ref{nlact}), however, is not symmetric 
under the general transformations due to the super-Weyl anomaly but 
it is invariant for the ones satisfying 
${\cal D}_-\vartheta_+=0$ as explained in Section 4 for $S_V^\pm$. 
Thus (\ref {nlact}) is invariant up to surface term under the 
super-M\"obius transformations with the coefficients being 
arbitrary function of $x^+$. This leads 
to the OSp(1,2) Kac-Moody (KM) symmetry, which can not be seen clearly 
in the local effective action (\ref{lcgfseff}). We can, however, 
extract the symmetry with the guideline 
of BRST invariance, to which we are to turn now.

In Appendix C we give the BRST transformations (\ref{slcbrst}) using 
the free ghost varaibles. Their consistency with the BRST invariance 
which are supposed to recover at the quantum level leads to 
\begin{equation}
\k\dl_-^3g_{++}=0~, \qquad \k\dl_-^2\chi_{++}=0 ~, \qquad
\k\dl_-\chi_+=0 ~, \label{sceq}
\end{equation}
from which the conservations of the gravitational stress tensor $T^g_{++}$ 
also follows. If we define conserved currents $J^a(x^+)$ 
and $\Psi^r(x^+)$ for $a=0,\pm$, $r=-1/2,~1/2$ by
\begin{eqnarray}
g_{++} &=& -\frac{1}{2\kappa}[J^{+}(x^{+})-2x^{-}
J^{0}(x^{+})+(x^{-})^2 J^{-}(x^{+})] ~,\nn\\
\chi_{++}&=& -\frac{1}{2\kappa}[\Psi^{-1/2}
(x^+)+x^-\Psi^{1/2}(x^+)] ~,\label{g++}
\end{eqnarray}
then they can be shown from their BRST transformation to satisfy OSp(1,2) 
KM algebra under the super-Poisson brackets. In our canonical 
approach the OSp(1,2) KM symmetry arises in this way. At first sight, 
our derivation might appear rather indirect and irrelevant to the 
invariance 
of the action. The reason for this is due to the fact that this symmetry 
is not present in (\ref{lcgfseff}) at the classical level in contrast 
with (\ref{nlact}). The BRST charge, however, can be regarded as the 
generator of the OSp(1,2) KM symmetry as noted in Appendix C, 
and the dynamics described by the action has been reflected to it through 
the equations of motion for such as the multiplier fields. 

Quantization can be acheived by replacing the super-Poisson brackets 
with the supercommutators times $-i$ for the fundamental brackets among the 
ghost variables and for the OSp(1,2) KM algebra. Free field operator 
ordering can be applied to the ghost variables to define operator products. 
As for the OSp(1,2) KM currents, we decompose them into lowering and rasing 
operators by the positive and negative frequency parts, by which we define 
operator ordering. The quantum mechanical stress tensor $T_{++}^g$ for 
gravity sector given in (\ref{stresstensor}) can be written in Sugawara 
form
\begin{equation}
T^g_{++}=-\frac{1}{\k'}:(\h_{ab}J^aJ^b+i\h_{rs}\Psi^r\Psi^s):
-\frac{1}{2}\dl_+J^0+\frac{i\k'}{8}
:\chi_+\dl_+\chi_+: ~, \label{qTg}
\end{equation}
where $\h_{ab}$ and $\h_{rs}$ are the inverse of the OSp(1,2) Killing 
metric. The parameter $\k'$ is given by
\begin{eqnarray}
\k'=\k-\frac{3}{8\p} ~. \label{kprime}
\end{eqnarray}
According to this, we have made a rescaling $\sqrt{\k}\chi_+
\rightarrow \sqrt{\k'}\chi_+$ in (\ref{qTg}). The stress tensor thus 
defined satisfies the Virasoro algebra with central charge 
\begin{equation}
c_g\equiv\frac{2k}{2k-3}+6k+\frac{1}{2} ~, \label{cg}
\end{equation}
where $k\equiv 4\p\k$ is the central charge of the OSp(1,2) KM current 
algebra. The second term in the rhs of (\ref{cg}) is present already 
at the classical level. The first term denotes the quantum Virasoro 
anomaly due to OSp(1,2) KM currents and the last one is due to $\chi_+$. 
The quantum theoretical BRST charge can be inferred from the classical 
expression (\ref{slcbrstcharg}), and is given by 
\begin{eqnarray}
Q_B&=&\int d\s:\Biggl[~c^+\biggl(T_{++}^X+T_{++}^{g}
+\frac{1}{2}T_{++}^{gh(2)}
+T_{++}^{gh(0)}+T_{++}^{gh(3/2)}+T_{++}^{gh(1/2)}\biggr) \nn\\
&&+c_+\biggl(-J^--ib\g_-\chi_++\frac{1}{8\k'}(\be_+\g_-)^2\biggr) 
\nn\\
&&+\g_-\biggl(-iJ^X_{++}-i\chi_+J^0+i\biggl(\k'+\frac{1}{8\p}\biggr)
\partial_+\chi_++2ib_{++}\g_-
-\frac{i}{\k'}\g_-(bJ^+-\be_+\Psi^{-1/2})\biggr) \nn \\
&&+\g_+\biggl(-4i\Psi^{1/2}+2ib\g_+
-\frac{i}{2}\be_+\g_-\chi_+\biggr)\Biggr]: ~.
\label{qslcbrstcharg}
\end{eqnarray}
This satisfies the nilpotency condition if and only if the 
centeral charge of the total stress tensors vanishes. We thus arrive 
at the KPZ condition \cite{kpz,pz,kura} for $N=1$ NSR superstring
\begin{equation}
c_{\rm tot}\equiv\frac{2k}{2k-3}+6k+\frac{1}{2}+\frac{3}{2}D-18=0 ~,
\label{totalcc}
\end{equation}
where the last two terms are the central charge of the 
string and the ghost sectors, respectively. 

\section{Summary and discussion}
In this paper we have investigated the $N=1$ RNS superstring at subcritical 
dimensions as 2D supergravity coupled to superconformal matter with 
special emphasis on the role played by the super-Virasoro and the 
super-Weyl 
anomaly. Naive action defined by free field ordering 
is not invariant under reparametrizations and local supersymmetry due to 
super-Virasoro anomaly but is under local Weyl rescalings and fermionic 
transformations. We have shown that the super-Virasoro anomaly can be 
canceled 
by introducing an additional term to the string action, which was also 
invariant under the local Weyl and fermionic transformations by 
construction. It was, however, nonlocal in the graviton and gravitino 
variables. The covariant nonlocal super-Liouville action with appropiate 
coefficients just has been shown to cancel the nonlocality and to recover 
the super-Weyl anomaly as expected from the perturvative analysis. 
This has lead to local but noncovariant super-Lioville action for the 2D 
supergravity, which can be interpreted as WZW term converting the 
super-Virasoro anomaly to the super-Weyl anomaly. 

We have also studied 
the quantization of 2D supergravity based on the super-Liouville action 
within the framework of BRST formalism and examined the BRST invariance 
in the superconformal gauge and in the light-cone gauge. 
In refs. \cite{d,dk,dhk} the key observation leading to the 
(super-)Liouville 
action is that the resulting effective theory should be (super)conformally 
invariant. Since the (super)conformal invariance is equivalent to the 
BRST invariance in (super)conformal gauge, our approach naturally 
reproduces the basic resutls of refs. \cite{d,dk,dhk} as argued in refs. 
\cite{fik1,fir-kt}. This can also be regarded as a canonical verification 
of the the functional measure ansatz of refs. \cite{d,dk,dhk}. We have also 
noted the subtlety concerning the cosmological term, which is not seen for 
the bosonic case \cite{ct}. It merely implies an artifact 
coming from the operator definition and could be resolved in principle in 
(super)conformally invariant formulation. 

In the case of light-cone gauge the residual OSp(1,2) KM symmetry 
can be extracted in a rather different way compared with the approaches 
of refs. \cite{gx,pz,aaz}, where the effective action obtained by 
integrating out the string variables possesses manifestly the symmetry. 
It is, however, hidden in our effective action and is extracted as a 
result of BRST invariance. One advantageous point of our approach is 
that we can investigate local properties of the theory instead of 
referring to the nonlocal effective action. 

We must reformulate our analysis for strings of finite length before 
the detailed comparision with the involved BRST analysis \cite{kura,ih} 
and the computations of physical quantities such as mass spectrum and 
string amplitudes, which are beyond the scope of the present paper. 

\vskip .5cm
We would like to thank H. Suzuki and Y. Igarashi for useful 
conversations. T. F. is supported in part by the Grant-in-Aid for 
Scientific Research from the Ministry of Education, Science 
and Culture under contract number 07740200.
\vspace{0.5cm}

\appendix
\section{BRST transformations}
\setcounter{equation}{0}

In this appendix we summarize the BRST transformations in 
covariant notation. The complete BRST transformations are
also given in the appendix of ref.\cite{fir-kt} with slightly different 
notation, where the scalar super-Liouville and super-Weyl ghost 
multiplets are included. In the present approach, these additional 
fields can be simply ignored since we do not impose the super-Weyl 
invariance at the quantum level. 

Let us introduce the reparametrization 
ghosts $C^\al$ $(\al=0,1)$, the local Lorentz ghost $C_L$ and 
the spinor ghost $\w$ for the local supersymmetry. Then the 
BRST transformations are given by
\begin{eqnarray}
&& \de X=C^\al\partial_\al X+\overline\w\psi ~, \nn \\
&& \de \psi=-\frac{1}{4}C_W\psi+C^\al\partial_\al\psi+\frac{1}{2}
C_L\r_5\psi
-i\r^\al\w(\partial_\al X-\overline\chi_\al\psi)+\w F_X ~, \nn \\
&& \de F_X=-\frac{1}{2}C_WF_X+C^\al\partial_\al F_X 
-i\overline\w\r^\al
[\nabla_\al\psi+i\r^\be(\partial_\be X-\overline\chi_\be\psi)\chi_\al
-\chi_\al F_X] ~,\nn \\
&& \de e_\al{}^a=C^\be\partial_\be e_\al{}^a+\partial_\al 
C^\be e_\be{}^a
+\e^a{}_bC_Le_\al{}^b-2i\overline\w\r^a\chi_\al~,
\nn \\
&& \de\chi_\al=C^\be\partial_\be\chi_\al
+\partial_\al C^\be\chi_\be+\frac{1}{2}C_L\r_5\chi_\al+\nabla_\al\w
-\frac{i}{4}\r_\al\w A ~, \nn \\
&& \de A=C^\al\partial_\al A
-4\e^{\al\be}\overline\w\r_5\nabla_\al\chi_\be
+i\overline\w\r^\al\chi_\al A ~, \nn \\
&& \de C^\al=C^\be\partial_\be C^\al+i\overline\w\r^\al\w ~, \nn \\
&& \de\w=C^\al\partial_\al\w+\frac{1}{2}C_L\r_5\w
-i\overline\w\r^\al\w\chi_\al ~, \nn \\
&& \de C_L=C^\al\dl_\al C_L+\frac{1}{2}A\overline\w\r_5\w
-i\e_{ab}e^{\be a}\overline\w\r^\al\w(\dl_\al e_\be{}^b
-\dl_\be e_\al{}^b-i\overline\chi_\al\r^b\chi_\be) ~, \label{cbrsttr}
\end{eqnarray}
where $F_X^\mu$ $(\mu=0,1,\cdots,D-1)$ and $A$ are, respectively, the 
auxiliary fields for the string and zweibein supermultiplets introduced 
to ensure the off-shell nilpotency of the BRST transformations. 
We should supplement the transformations for the anti-ghosts and 
multiplier fields, which have been omitted here.

\section{Derivation of super-Virasoro algebra}
\setcounter{equation}{0}

In this appendix we describe operator ordering of the super-Virasoro 
generator for the super-Liouville field and their commutation relations 
in the presence of the cosmological term. 

We first note that the gravitational super-Viraosoro generators can be 
defined in the manner similar to (\ref{sVconst 2}) as 
\begin{equation}
\Phi_\pm=-\frac{\de S_g}{\de \lam^\pm}~,\qquad
{\cal I}_\pm=\mp i\frac{\de S_g}{\de \nu_\mp}~. 
\label{gsVconst}
\end{equation}
When expressed in terms of the canonical variables, $\xi$ and its conjugate 
momentum, for instance, these can be seen to satisfy super-Virasoro algebra 
with central charge $24\p\k$ under the super-Poisson brackets. This 
situation is not changed when the contribution from the cosmological term 
are included.

For notational simplicity, we use here the rescaled variables $\vp=\g\xi$, 
$\chi_\mp=\g\Lam\mp$ and 
$f=\g a$ with $\displaystyle{\g\equiv\sqrt{\frac{\k}{2}}}$ for the action
(\ref{qSL}). We make also an appropriate translation of $\vp$ by a 
constant to keep fixed the cosmological constant $\m$. Then the 
super-Virasoro generators for the gravitational sector including the 
contributions from the cosmological term can be written as 
\begin{eqnarray}
\Phi_\pm&=&\frac{1}{4}\Th_\pm^2-\g\Th_\pm'\pm \frac{i}{2}\chi_\pm\chi'_\pm
+\m^2e^{2\be\vp}+i\m\be e^{\be\vp}\chi_+\chi_- ~,\nn \\
{\cal I}_\pm&=&\pm\chi_\pm\Th_\pm\mp4\g\chi_\pm'\mp2\m e^{\be\vp}\chi_\mp ~,
\label{sValg}
\end{eqnarray}
where $\displaystyle{\be\equiv\frac{\al}{\g}}$ and 
$\Th_\pm\equiv\vp'\pm\p_\vp$ with $\p_\vp=\dot\vp$ being the canonical 
momentum conjugate to $\vp$. We have inserted a parameter $\al$ to take 
quantum corrections into account. For the classical value 
$\displaystyle{\be\g=\frac{1}{2}}$, they coincide with $T^L_{\pm\pm}$ 
and $J^L_{\pm\pm}$ given in (\ref{stsc}). 

In quantum theory, we assume equal-time supercommutation relations 
\begin{eqnarray}
[\vp(\s),\p_\vp(\s')]=i\de(\s-\s'), \qquad [\chi_\pm(\s),\chi_\pm(\s')]
=\de(\s-\s') ~. \label{ETC}
\end{eqnarray}
To define operator ordering, we introduce harmonic oscillators $a_k$ and 
$b_k$ by 
\begin{eqnarray}
\vp(\s)=\frac{i}{2\sqrt{\p}}\int_{-\infty}^{+\infty}
\frac{dk}{k}(a_ke^{-ik\s}+b_ke^{ik\s})~,~~
\p_\vp(\s)=\frac{1}{2\sqrt{\p}}\int_{-\infty}^{+\infty}
dk(a_ke^{-ik\s}+b_ke^{ik\s})~.
\end{eqnarray}
We do not care for the infrared singularity due to the infinite region 
for $\s$ since most of our arguments are irrelevant to this problem and 
it can be avoided by considering finite interval for $\s$. These 
oscillators satisfy 
\begin{eqnarray}
[a_k,a_{k'}]=[b_k,b_{k'}]=k\de(k-k')~, \qquad [a_k,b_{k'}]=0 ~. \label{ho}
\end{eqnarray}
We then define $a_k$ and $b_k$ as lowering operators and $a_k^\dagger=
a_{-k}$ and $b_k^\dagger=b_{-k}$ as rasing operators for $k>0$. We use 
similar decomposition for $\chi_\pm$. The lowering and rasing parts of 
$\Th_\pm$ and $\chi_\pm$ can be written in the manner similar to 
(\ref{creanniop}) as
\begin{eqnarray}
&&\Th_+^{(\pm)}(\s)=\int d\s'\de^{(\mp)}(\s-\s')\Th_+(\s')~, \qquad
\Th_-^{(\pm)}(\s)=\int d\s'\de^{(\pm)}(\s-\s')\Th_-(\s')~, \nn \\
&&\chi_+^{(\pm)}(\s)=\int d\s'\de^{(\mp)}(\s-\s')\chi_+(\s')~, \qquad
\chi_-^{(\pm)}(\s)=\int d\s'\de^{(\pm)}(\s-\s')\chi_-(\s')~. \label{opdecomp}
\end{eqnarray}
This define an operator ordering. 

Let us denote by $\Phi^{(0)}_\pm$ and ${\cal I}^{(0)}_\pm$ the 
super-Virasoro 
operators for $\m=0$ defined by the normal ordering, then they satisfy 
super-Virasoro algebra with central charge 
$\displaystyle{\frac{3}{2}+24\p\k}$. The deviation from the classical 
value is due to quantum fluctuations of the super-Liouville fields 
themselves. 

We next define vertex-like operators by
\begin{eqnarray}
V_\be=:\exp \be\vp: ~, \quad 
V_{\be\pm}=:\exp \be\vp\chi_\pm: ~, \quad
V_{\be+-}=:\exp\be\vp\chi_+\chi_-: ~,
\label{vertexop}
\end{eqnarray}
and examine the commutation relations with $\Phi^{(0)}_\pm$ and 
${\cal I}^{(0)}_\pm$. We find 
\begin{eqnarray}
[\Phi^{(0)}_\pm(\s),V_\be(\s')]&=&\mp\frac{i}{2}
\be:\Th_\pm V_\be:\de(\s-\s')
\pm i\Bigl(\be\g-\frac{\be^2}{8\p}\Bigr)\partial_\s(V_\be\de(\s-\s')) ~, 
\nn \\
~[\Phi^{(0)}_\pm(\s),V_{\be\mp}(\s')]&=&\mp\frac{i}{2}\be:\Th_\pm 
V_{\be\mp}:\de(\s-\s')\pm i\Bigl(\be\g-\frac{\be^2}{8\p}\Bigr) 
\partial_\s(V_{\be\mp}\de(\s-\s')) ~, \nn \\
~[\Phi^{(0)}_\pm(\s),V_{\be\pm}(\s')]&=&\mp i
:(\frac{\be}{2}\Th_\pm V_{\be\pm}
+V_\be\partial_\s\chi_\pm):\de(\s-\s') \nn \\
&& \pm i\Bigl(\be\g-\frac{\be^2}{8\p}+\frac{1}{2}\Bigr) 
\partial_\s(V_{\be\pm}\de(\s-\s')) ~, \nn \\
~[\Phi^{(0)}_\pm(\s),V_{\be+-}(\s')]&=&\mp i
:(\frac{\be}{2}\Th_\pm V_{\be+-}
+V_\be\partial_\s\chi_+\chi_-):\de(\s-\s') \nn \\
&& \pm i\Bigl(\be\g-\frac{\be^2}{8\p}+\frac{1}{2}\Bigr) 
\partial_\s(V_{\be+-}\de(\s-\s')) ~, \nn \\
~[{\cal I}^{(0)}_\pm(\s),V_\be(\s')]&=&-i\be V_{\be\pm}\de(\s-\s')~, \nn \\
~[{\cal I}^{(0)}_\pm(\s),V_{\be\mp}(\s')]&=&
\mp i\be V_{\be+-}\de(\s-\s')~, \nn \\
~[{\cal I}^{(0)}_\pm(\s),V_{\be\pm}(\s')]&=&\pm:\Th_\pm V_\be:\de(\s-\s')
\mp\Bigl(4\g-\frac{\be}{2\p}\Bigr)\partial_\s(V_{\be\mp}\de(\s-\s'))~, 
\nn \\
~[{\cal I}^{(0)}_\pm(\s),V_{\be+-}(\s')]
&=&:\Th_\pm V_{\be\mp}:\de(\s-\s')
-\Bigl(4\g-\frac{\be}{2\p}\Bigr)\partial_\s(V_{\be\mp}\de(\s-\s'))~. 
\label{precomu}
\end{eqnarray}
The operators (\ref{vertexop}) themselves satisfy
\begin{eqnarray}
&&[V_\be(\s),V_\be(\s')]=[V_\be(\s),V_{\be\pm}(\s')]
=[V_\be(\s),V_{\be+-}(\s')] \nn \\
&&=[V_{\be+}(\s),V_{\be-}(\s')]
=[V_{\be+-}(\s),V_{\be+-}(\s')]=0 ~, \label{vcom}
\end{eqnarray}
where we have used the relation
\begin{eqnarray}
&& V_\be(\s)V_\be(\s')=:V_\be(\s)V_\be(\s'):
\exp\Biggl[\frac{\be^2}{2\p}\sum_{n=1}^{\infty}
\frac{1}{n}\cos\frac{2\p n}{L}(\s-\s')\Biggr] ~, \nn \\
&& [:\chi_+\chi_-(\s):, :\chi_+\chi_-(\s'):]=0 ~,
\end{eqnarray}
with $L$ being the length of the interval for $\s$ to avoid infrared 
divergence. The commutors among $V_{\be\pm}$ and $V_{\be+-}$ are problematic
since they contain a divergent quantity $V_\be^2$, which can not be removed 
by the simple ordering prescription. This situation also occurs even 
in free theories. We do not pursue here the handling of the divergence 
of this type. Instead, we simply assume the existence of suitable way 
of defining operators like $V_\be^2$ consistently with the superconformal 
tarnsformations and content ourselves to write the commutators as 
\begin{eqnarray}
[V_{\be\pm}(\s),V_{\be\pm}(\s')]
=V_\be^2\de(\s-\s')~\qquad 
[V_{\be\pm}(\s),V_{\be+-}(\s')]
=\pm V_\be^2\chi_\mp\de(\s-\s')~. \label{vp2}
\end{eqnarray}
We now define quantum super-Virasoro generators by
\begin{eqnarray}
\Phi_\pm=\Phi_\pm^{(0)}+\m^2V_\be^2+i\m\be V_{\be+-} ~, \qquad
{\cal I}_\pm={\cal I}_\pm^{(0)}\mp2\m V_{\be\mp} \label{qsVop}
\end{eqnarray}
Using (\ref{precomu}) and (\ref{vcom}), one can show the following 
commutation relations 
\begin{eqnarray}
[\Phi_\pm(\s),\Phi_\pm(\s')]&=&\pm i(\Phi_\pm(\s)+\Phi_\pm(\s'))
\partial_\s\de(\s-\s')\mp i\Biggl(2\g^2+\frac{1}{16\p}\Biggr)
\partial_\s^3\de(\s-\s') \nn \\
&&\pm 2i\Biggl(\be\g-\frac{\be^2}{8\p}-\frac{1}{2}\Biggr)(
\m^2V_\be^2(\s)+i\m\be V_{\be+-}(\s)+(\s\rightarrow\s'))
\partial_\s\de(\s-\s') ~, \nn \\
~[\Phi_\pm(\s),{\cal I}_\pm(\s')]&=&\pm\frac{3}{2}i{\cal I}_\pm(\s)
\partial_\s\de(\s-\s')\pm\frac{i}{2}\partial_\s{\cal I}_\pm\de(\s-\s') \nn \\
&&-4i\m\Biggl(\be\g-\frac{\be^2}{8\p}-\frac{1}{2}\Biggr)
\Biggl(\frac{3}{2}V_{\be\mp}(\s)\partial_\s\de(\s-\s')
+\frac{1}{2}\partial_\s V_{\be\mp}\de(\s-\s')\Biggr) ~,\nn \\
~[{\cal I}_\pm(\s),{\cal I}_\pm(\s')]&=&4\Phi_\pm\de(\s-\s')
-8\Biggl(2\g^2+\frac{1}{16\p}\Biggr)\partial_\s^2\de(\s-\s') ~,\nn \\
~[\Phi_+(\s),\Phi_-(\s')]&=&2i\Biggl(\be\g-\frac{\be^2}{8\p}
-\frac{1}{2}\Biggr)\partial_\s(\m^2V_\be^2+i\m\be V_{\be+-})
\de(\s-\s') ~, \nn \\
~[\Phi_\pm(\s),{\cal I}_\mp(\s')]&=&-2i\m
\Biggl(\be\g-\frac{\be^2}{8\p}-\frac{1}{2}\Biggr)
(V_{\be\pm}(\s)\partial_\s\de(\s-\s')
-\partial_\s V_{\be\pm}\de(\s-\s'))~, \nn \\
~[{\cal I}_+(\s),{\cal I}_-(\s')]&=&-\frac{8\m}{\be}
\Biggl(\be\g-\frac{\be^2}{8\p}-\frac{1}{2}\Biggr)
\partial_\s V_\be\de(\s-\s') ~. \label{PIcom}
\end{eqnarray}
We see that the contribution from the cosmological term does not 
shift the central charge and (\ref{qsVop}) fulfills super-Virasoro 
algebra for $\be$ satisfying
\begin{eqnarray} 
\be\g-\frac{\be^2}{8\p}=\frac{1}{2} ~. \label{beeq}
\end{eqnarray}
This combined with (\ref{kappa}) gives (\ref{alpha}). 

\section{Light-cone gauge-fixing}
\setcounter{equation}{0}

We shall argue the light-cone gauge-fixing in some detail. The ghost 
and gauge-fixing part of the action is give by 
\begin{eqnarray}
S_{\rm gh}&=&\int d^2x(-\overline C_{-+}\de e_+{}^+
-\overline C_{++}\de e_-{}^+-\overline C_{+-}\de e_-{}^-
+\overline \be_{+-}\de\chi_{-+}+\overline \be_{++}\de\chi_{--}) ~,\nn \\
S_{\rm GF}&=&\int d^2x\Bigl[-B_{-+}(e_+{}^+-1)-B_{++}e_-{}^+-B_{+-}
(e_-{}^--1) \nn \\
&& -\z_{+-}\chi_{-+}-\z_{++}\chi_{--}\Bigr] ~, \label{ghGF}
\end{eqnarray}
where $\overline C_{\al a}$, $\overline\be_{+\pm}$ are the anti-ghosts 
and $B_{\al a}$, $\z_{+\pm}$ are the multiplier fields for the gauge 
conditions (\ref{lcgc}). They satisfy the BRST transformations 
\begin{eqnarray}
\de\overline C_{\al a}=-B_{\al a} ~, \qquad 
\de\overline\be_{+\pm}=-\z_{+\pm} ~, \qquad 
\de B_{\al a}=0 ~, \qquad \de\z_{+\pm}=0 .
\end{eqnarray}
The BRST transformations of the zweibeins and gravitinos are given 
in (\ref{cbrsttr}). To find the BRST transformations of the anti-ghost 
variables, we need the multiplier fields, which can be found by taking 
variations of the total action with respect to the variables fixed by 
the gauge conditions and then imposing the light-cone gauge. For instance, 
the $\z_{+-}$ and $B_{+-}$ can be obtained as follows; 
\begin{eqnarray}
\z_{+-}&=&\frac{2}{\sqrt{-e_1{}^-}}\frac{\de S_{\rm GF}}{\de\n_-}
\Biggr|_{\rm l.c.g.} \nn\\
&=&-\frac{2}{\sqrt{-e_1{}^-}}\frac{\de}{\de\n_-}(S_X+S_g
+S_{\rm cosm}+S_{\rm gh})
\Biggr|_{\rm l.c.g.} \nn \\
&=&\frac{2i}{\sqrt{-e_1{}^-}}\Biggl({\cal J}_-+{\cal I}_-
+i\frac{\de S_{\rm gh}}{\de\n_-}\Biggr)\Biggr|_{\rm l.c.g.} \nn\\
&=&8i\k\dl_-\chi_{++}+4i\overline C_{+-}\w_+
-\frac{1}{2}\dl_+\overline\be_{+-}C^+-\frac{1}{4}\overline\be_{+-}
\dl_+C^++2i\overline\be_{+-}\w_-\chi_{+-} ~,\nn\\
B_{+-}&=&\frac{2}{e_1{}^-}\Biggl(
\frac{\de S_{\rm GF}}{\de\lam^-}-\frac{1}{8}\sqrt{-e_1{}^-}
\z_{+-}\Lam_-\Biggl)\Biggr|_{\rm l.c.g.} \nn\\
&=&\frac{2}{e_1{}^-}\Biggl(-\frac{\de}{\de\lam^-}(S_X+S_g+S_{\rm cosm}
+S_{\rm gh})+\frac{i}{4}\Lam_-\Biggl({\cal J}_-+{\cal I}_-
+i\frac{\de S_{\rm gh}}{\de\n_+}\Biggr)\Biggr)\Biggr|_{\rm l.c.g.} \nn\\
&=&\frac{2}{e_1{}^-}\Biggl(\vp_-+\Phi_-
+\frac{i}{4}\Lam_-({\cal J}_-+{\cal I}_-)-\frac{\k}{8}g_{11}A^2
-\frac{\de S_{\rm gh}}{\de\lam^-}+\frac{i}{4}\Lam_-
\frac{\de S_{\rm gh}}{\de\n_+}\Biggr)\Biggr|_{\rm l.c.g.} \nn\\
&=&-\frac{\k}{2}\dl_-^2g_{++}+\frac{1}{2}\dl_+\overline C_{+-}C^+
+2i\overline C_{+-}\w_-\chi_{+-}-\frac{1}{4}\overline\be_{+-}\w_-A
+\frac{\k}{4}A^2 ~,
\label{zpm}
\end{eqnarray}
where $|_{\rm l.c.g.}$ denotes restriction to the light-cone gauge and 
$\Phi_-$, ${\cal I}_-$ are defined in (\ref{gsVconst}). In derivng 
these results, use has been made of the equaitons of motion for 
$g_{++}$ and $\chi_{++}$, i.e., 
\begin{eqnarray}
\frac{\k}{4}\dl_-^2g_{++}&=&\frac{1}{g_{11}}
\Biggl((\vp_-+\Phi_-)\Biggr|_{\rm l.c.g.}+4\k\chi_{++}
\dl_-\chi_{++}\Biggr), \nn\\
4\k\dl_-\chi_{++}&=&\frac{1}{\sqrt{g_{11}}}({\cal J}_-
+{\cal I}_-)\Biggr|_{\rm l.c.g.} . \label{gppeom}
\end{eqnarray}
The rest of the multipliers can be found in the same way.

We now turn to the ghost action in (\ref{ghGF}). In the light-cone gauge 
this reduces to 
\begin{eqnarray}
S_{\rm gh}&=&\int d^2x\Biggl[ -\frac{1}{2}(\overline C_{++}-g_{++}\overline 
C_{+-}+\overline\be_{+-}\chi_{++}+\overline\be_{++}\chi_{+-})\dl_-C^+ \nn\\
&&-\frac{1}{2}\overline C_{+-}(\dl_+C^++\dl_-C^--8i\w_-\chi_{+-}) \nn\\
&&+\frac{1}{2}\overline\be_{+-}(\dl_-\w_+-\frac{1}{2}\w_-A)+\frac{1}{2}
\overline\be_{++}\dl_-\w_-\Biggr] ~, \label{lcghact}
\end{eqnarray}
where $\overline C_{-+}$ and $C_L$ have been eliminated by virtue of
their equations of motion. From (\ref{lcghact}) we see that the ghost 
variables satisfy 
\begin{eqnarray}
\dl_-C^+=0 ~, \qquad 
\dl_-^3C^-=0~, \qquad \dl_-\w_-=0 , \qquad \dl_-^2\w_+=0 ~ \label{osp21}
\end{eqnarray}
in accord with the residual OSp(1,2) KM symmetry of the effective 
action. In deriving these use has been made of the equations of motion 
for $\chi_{+-}$ and $A$. Then the BRST charge can be regarded as 
the infinitesimal generator of this symmetry. 

Eqs.(\ref{osp21}) suggest that all the ghost variables can be converted 
to free fields. This is implemented by the field redefinitions
\begin{eqnarray}
C^+&=&c^+ ~,\nn\\
C^-&=&c_+-\frac{x^-}{2}\dl_+c^++ix^-\g_-\chi_++\frac{i}{\k}(x^-)^2b
\g^2_- ~,\nn\\
\w_+&=&\g_++\frac{x^-}{4\k}\be_+\g_-^2 ~, \nn\\
\w_-&=&\g_- ~,\nn\\
\overline C_{++}&=&2\Biggl\{b_{++}-\frac{x^-}{2}\dl_+b+g_{++}b
-\be_+\chi_{++}
-\frac{1}{4}\be_{++}\chi_+ \nn\\
&&-\frac{x^-}{2\k}b\be_{++}\g_-
+\frac{x^-}{16\k}\be_+^2\g_-\chi_++\frac{(x^-)^2}{8\k^2}b\be_+^2\g_-^2
\Biggr\} ~,\nn\\
\overline C_{+-}&=&2b ,\nn\\
\overline\z_{++}&=&2\Biggl\{\be_{++}+ix^-b\chi_+
-\frac{x^-}{4\k}\be^2_+\g_-\Biggr\} ~,\nn\\
\overline\z_{+-}&=&2\be_+ ~,\nn\\
\chi_{+-}&=&\frac{1}{4}\chi_++\frac{x^-}{2\k}b\g_- ~.\label{freegh}
\end{eqnarray}
We thus arrive at the ghost part of the effective action (\ref{lcgfseff}). 

It is straightforward to rewrite the BRST transformations in terms of 
these free ghost variables. They are give by
\begin{eqnarray}
\de g_{++}&=&\frac{1}{2}c^+\dl_+ g_{++}+g_{++}\dl_+ c^+
+\frac{1}{2}\Bigl(c_+-\frac{x^-}{2}\dl_+ c^+\Bigr)\dl_- g_{++}
-\frac{1}{2}\dl_+\Bigl(c_+-\frac{x^-}{2}\dl_+ c^+\Bigr) \nn \\
&& +4i\g_+\chi_{++}-i\g_-\chi_+\Bigl(1-\frac{x^-}{2}\dl_-\Bigr)g_{++}
-\frac{i}{2}x^-\dl_+(\g_-\chi_+) \nn \\
&& +\frac{i}{\k}x^-\be_+\g_-^2\chi_{++}
-\frac{2i}{\k}b\g_-^2\Bigl(x^--\frac{(x^-)^2}{4}\dl_-\Bigr)g_{++}
-\frac{i}{2\k}(x^-)^2\dl_+(b\g_-^2) ~, \nn \\
\de\chi_{++}&=&\frac{1}{2}c^+\dl_+\chi_{++}+\frac{3}{4}\dl_+ c^+\chi_{++}
+\frac{1}{2}\Bigl(c_+-\frac{x^-}{2}\dl_+\Bigr)\dl_-\chi_{++}
+\frac{1}{2}\dl_+\g_+-\frac{1}{4}\g_+\dl_- g_{++} \nn \\
&& -\frac{i}{2}\g_-\chi_+(1-x^-\dl_-)\chi_{++}
-\frac{i}{\k}b\g_-^2\Bigl\{x^--\frac{(x^-)^2}{2}\dl_-\Bigr\}\chi_{++} \nn \\
&& +\frac{1}{4\k}\be_+\g_-^2
\Bigl(1-\frac{x^-}{4}\dl_-\Bigr)g_{++}
+\frac{x^-}{8\k}\dl_+(\be_+\g_-^2) ~, \nn \\
\de \chi_+&=&\frac{1}{2}c^+\dl_+\chi_+-\frac{1}{4}\dl_+ c^+\chi_+
+2\dl_+\g_- \nn \\
&& +\g_-\Bigl(1-\frac{x^-}{2}\dl_-\Bigr)\dl_- g_{++}
-\frac{2}{\k}\g_-\Bigl(bc_+-\frac{1}{2}\be_+\g_+\Bigr) ~, \nn \\
\de c^+&=&\frac{1}{2}c^+\dl_+ c^++2i\g_-^2 ~, \nn \\
\de c_+&=&\frac{1}{2}c^+\dl_+ c_++\frac{1}{2}\dl_+ c^+c_++2i\g_+^2 ~, \nn \\
\de\g_-&=&\frac{1}{2}c^+\dl_+\g_--\frac{1}{4}\dl_+ c^+\g_- ~, \nn \\
\de\g_+&=&\frac{1}{2}c^+\dl_+\g_++\frac{1}{4}\dl_+ c^+\g_+
-\frac{i}{2}\g_+\g_-\chi_+  \nn \\
&&-2i\g_-^2\Bigl(1-\frac{x^-}{2}\dl_-\Bigr)\chi_{++}
+\frac{1}{4\k}c_+\be_+\g^2_- ~,\nn \\
\de b_{++}&=& 
T_{++}^{X}+T_{++}^{g}+\frac{1}{2}T_{++}^{gh(2)}+T_{++}^{gh(0)}
+T_{++}^{gh(3/2)}
+T_{++}^{gh(1/2)} ~,
\nn\\
\de b&=&\frac{\k}{4}\dl_-^2g_{++}-\frac{1}{2}\dl_+ bc^+
-ib\g_-\chi_++\frac{1}{8\k}(\be_+\g_-)^2 ~, \nn \\
\de \be_{++}&=&iJ^X_{++}+\frac{i\k}{2}\chi_+
\Bigl(1-\frac{x^-}{2}\dl_-\Bigr)\dl_- g_{++}-i\k\dl_+\chi_+ \nn \\
&& +\frac{1}{2}\dl_+\be_{++}c^++\frac{3}{4}\be_{++}\dl_+ c^+
-4ib_{++}\g_- \nn \\
&& -4ib\g_-\Biggl\{1-\frac{x^-}{2}\dl_-+\frac{(x^-)^2}{8}\dl_-^2
\Biggr\}g_{++}
+4i\be_+\g_-\Bigl(1-\frac{x^-}{2}\dl_-\Bigr)\chi_{++} \nn \\
&& -i\bigl(bc_+-\frac{1}{2}\be_+\g_+\bigr)\chi_+
-\frac{1}{4\k}\be_+^2\g_-c_+ ~, \nn \\
\de\be_+&=&-4i\k\dl_-\chi_{++}+\frac{1}{2}\dl_+\be c^+
+\frac{1}{4}\be_+\dl_+ c^+-4ib\g_++\frac{i}{2}\be_+\g_-\chi_+ ~, 
\label{slcbrst}
\end{eqnarray}
where the stress tensors and supercurrents are given by 
\begin{eqnarray}
T_{++}^X&\equiv& \vp_+
=\frac{1}{4}[(\dl_+X+g_{++}\dl_-X)^2+i\psi_+\dl_+\psi_+] ,\nn\\
J_{++}^X&\equiv & \CJ_+=\psi_+(\dl_+X+g_{++}\dl_-X) ~, \nn \\
T_{++}^{g}&\equiv& \frac{\k}{2}\biggl[~\frac{1}{4}(\dl_-g_{++})^2
-\frac{1}{2}g_{++}\dl_-^2g_{++}
-\frac{1}{2}(\dl_+-\frac{x^-}{2}\dl_-\dl_+)\dl_-g_{++}\biggr]  \nn\\
& & +4i\k\chi_{++}\dl_-\chi_{++}+\frac{i\k}{8}\chi_+\dl_+\chi_+ ~, \nn \\
T_{++}^{gh(2)}&\equiv & -\frac{1}{2}\dl_+b_{++}c^+-b_{++}\dl_+c^+ ~,
\nn\\
T_{++}^{gh(0)}&\equiv& \frac{1}{2}\dl_+bc_+ ~, \nn \\
T_{++}^{gh(3/2)}&\equiv &\frac{3}{4}\be_{++}\dl_+\g_-
+\frac{1}{4}\dl_+\be_{++}\g_- ~, \nn \\
T_{++}^{gh(1/2)}&\equiv&\frac{1}{4}\be_+\dl_+\g_+
-\frac{1}{4}\dl_+\be_+\g_+ ~. \label{stresstensor}
\end{eqnarray}
In (\ref{slcbrst}) the transformations for string variables have 
been suppressed. Since the ghost variables are free fields, we can 
unambiguously find the classcial BRST charge from the BRST 
transformations of the ghost variables as 
\begin{eqnarray}
Q_B&=&\int d\s\Biggl[~c^+\biggl(T_{++}^X+T_{++}^{g}+
\frac{1}{2}T_{++}^{gh(2)}
+T_{++}^{gh(0)}+T_{++}^{gh(3/2)}+T_{++}^{gh(1/2)}\biggr) \nn\\
&&+c_+\biggl(\frac{\k}{4}\dl_-^2g_{++}-ib\g_-\chi_+
+\frac{1}{8\k }(\be_+\g_-)^2\biggr) \nn\\
&&+\g_-\biggl(-iJ^X_{++}
-\frac{i\k}{2}\chi_+\biggl(1-\frac{x^-}{2}\dl_-\biggr)\dl_-g_{++}
+i\k\partial_+\chi_++2ib_{++}\g_- \nn \\
&& +2ib\g_-\biggl(1-\frac{x^-}{2}\dl_-
+\frac{(x^-)^2}{8}\dl^2_-\biggr)g_{++} 
-2i\be_+\g_-\biggl(1-\frac{x^-}{2}\dl_-\biggr)\chi_{++}\biggr) \nn \\
&&+\g_+\biggl(4i\k\dl_-\chi_{++}+2ib\g_+
-\frac{i}{2}\be_+\g_-\chi_+\biggr)\Biggr] ~.
\label{slcbrstcharg}
\end{eqnarray}

\end{document}